\newcommand{\ep}{\varepsilon}
\newcommand{\nn}{\nonumber}

\newcommand{\ord}[2]{{\cal O}(#1 ^{#2})}

\newcommand{\SCR}[1]{{\mathscr #1}}

\newcommand{\CAL}[1]{{\cal #1}}

\newcommand{\MAT}[1]{\left(\begin{array}{cccccccccc}#1\end{array}\right)}
\newcommand{\J}[1]{\left\langle #1 \right\rangle}
\newcommand{\D}[1]{{\mathscr D}( #1 )}

\documentclass[preprint,12pt]{article}

\usepackage[top=30truemm, bottom=30truemm, left=25truemm, right=25truemm]{geometry}

\usepackage{fancyhdr}
\usepackage{amsthm}
\usepackage{amsmath,amssymb,latexsym,amsfonts,mathrsfs}

 \newtheorem{thm}{Theorem}[section]
 
 \newtheorem{Ass}[thm]{Assumption}
 \newtheorem{lem}[thm]{Lemma}
 \newtheorem{Cor}[thm]{Corollary}
 \newtheorem{prop}[thm]{Proposition}
 \theoremstyle{definition}
 \newtheorem{defn}[thm]{Definition}
 \newtheorem{rem}[thm]{Remark}
 
 \numberwithin{equation}{section}

\newcommand{\Proof}[2][Proof]{
\begin{proof}[{\bf #1}]
#2
\end{proof}
}

\usepackage{amssymb}

\begin{document}

\begin{flushleft}
{ \Large \bf Quantum scattering with time-decaying
 harmonic oscillators}
\end{flushleft}

\begin{flushleft}
{\large Masaki KAWAMOTO}\\
{
Department of Engineering for Production, Graduate School of Science and Engineering, Ehime University, 3 Bunkyo-cho Matsuyama, Ehime, 790-8577. Japan \\ 
Email: kawamoto.masaki.zs@ehime-u.ac.jp 
}
\end{flushleft}
\begin{abstract}

By controlling the coefficients and decaying order of time-decaying harmonic
 potentials, the velocity of a quantum particle can be
 reduced by the effect of the harmonic potentials, without trapping the particle. In this study, we have considered a
 quantum system with controlled harmonic potentials. By defining the wave
 operators with suitable ranges, we characterized the range of wave operators with respect to the
 short-range potentials, $V(t,x)$, satisfying $|V(t,x)| = o
 ((1+|x|)^{-1/(1- \lambda)})$ for some $\lambda$, $0 \leq  \lambda < 1/2$.
\end{abstract}

\begin{flushleft}


\end{flushleft}

\section{Introduction}

Herein, we have considered the dynamics of a quantum particle under the influence of
time-dependent harmonic potentials. We define the time-dependent harmonic
potentials by $k(t) x^2/2$, where $x=(x_1, \cdots , x_n)$,
$n \in {\bf N}$ is the position of the particle, and {{$k \in
L^{\infty} ({\bf R})$}}. The free time-dependent Hamiltonian is then described by
\begin{align}\label{1}
H_0(t) = p^2/(2m) + k(t)x^2/2,
\end{align}
where $p=(p_1,\cdots,p_n) = -i (\partial _1, \cdots ,\partial _n)$ and
$m>0$ are the momentum and mass of the particle,
respectively. 
 $U_0(t,s)$ is determined by the propagator for $H_0(t)$, i.e., a family of unitary operators $\{U_0(t,s)\}_{(t,s) \in {\bf R}^2}$
in $L^2({\bf R}^n)$,
with each element satisfying 
\begin{align*}
&i \partial _t U_0(t,s) =H_0(t)U_0(t,s) , \quad 
i \partial_s U_0(t,s) = - U_0(t,s) H_0(s), \\ 
&U_0(t,\theta)U_0(\theta ,s) =U_0(t,s), \quad U_0(s,s)= \mathrm{Id}. 
\end{align*} 
 The asymptotic behavior of the particle
governed by $H_0(t)$ is investigated here. The observables are defined as $$ x_0(t) := U_0(t,0)^{\ast} x
U_0(t,0), \quad p_0(t) := U_0(t,0)^{\ast} p
U_0(t,0)
$$, and from the commutator calculation
\begin{align*}
 & x' _0(t) = p_0(t) /m, \quad p_0(0) = p, \\ 
 & x''_0(t) + (k(t)/m) x_0(t) = 0, \quad x_0(0) = x , \ x'_0(0) = p/m,
\end{align*} 
where $x'_0(t) = (\partial _t x_0)(t) $ and $x''_0(t) = (\partial
_t^2 x_0)(t)$. We define the fundamental solutions $\zeta _1 (t)$ and $\zeta
_2 (t)$ as solutions to 
\begin{align}\label{2}
\zeta _j ''(t) + \left( 
\frac{k(t)}{m}
\right)\zeta _j (t) =0, \quad 
\begin{cases}
\zeta _1 (0) = 1, \\ 
\zeta _1 '(0) = 0,
\end{cases}
\quad 
\begin{cases}
\zeta _2 (0) = 0, \\ 
\zeta _2 '(0) = 1. 
\end{cases}
\end{align}
Therefore, $x_0(t) = \zeta _1 (t) x + \zeta _2 (t) p /m$ and $p_0 (t) = m \zeta _1 '(t)x + \zeta _2 '(t) p $. If $k(t)$ is a periodic function with respect to $t$, the equations in \eqref{2}
are called {\em Hill's equations}, and the asymptotic behavior of the
solutions of \eqref{2} are well known.

 Korotyaev \cite{Ko};
Hagedorn, Loss and Slawny \cite{HLS}; Huang \cite{Hu}; Adachi and Kawamoto
\cite{AK2}; and Kawamoto \cite{Ka} considered 
Hill's equation, spectral theory, scattering theory, and the
associated issues. 

The case where $k(t)$ decays with
time in \eqref{2} can now be considered. If $k(t) = c_0 t^{-2-c_1}$ for $c_0 \neq 0$ and $c_1
>0$, the solution of \eqref{2} satisfies $\zeta
_1 (t)/t \sim c_2  $ and $\zeta _2 (t)/t \sim c_3 $ for some
$c_2  ,c_3  \in {\bf R}$, e.g., see Willett \cite{W} and Naito
\cite{Na}. Hence, the classical trajectory is 
similar to the $k(t) \equiv 0$ case. Although this system is 
similar to a system considered by Yafaev \cite{Yaf}; Kitada and Yajima
\cite{KY}; and Kato and Yoneyama \cite{Kyo}, scattering theory for this case, to the best of our knowledge, has not been considered
thus far. 

For the case where $k(t) =
c_4 t^{-2 + c_5}$, $c_4 \neq 0$, and $c_5 >0$, the solution of \eqref{2}
behaves like $\zeta _1 (t) \sim t^{c_6} \cos (\theta _1 (t)) $ and $\zeta
_2 (t) \sim  t^{c_6} \sin (\theta _2 (t))$, which is asymptotic in $t$ with $0 < c_6 <1$ and some
functions, such as $\theta _1
(t)$ and $\theta _2 (t) $, satisfying $|\theta _j (t)| \to \infty $ as $t
\to \infty$. This can be demonstrated by applying the Hochstadt approach \cite{Ho}. In this case, the quantum
particle not only moves out of any compact regions, but also returns to the
origin infinitely many times. This movement makes it difficult to prove
the existence of wave operators; hence there are no works associated
with scattering theory for this model.

Hence, in this study, we consider a simple model of time-decaying harmonic oscillators,  $k(t)$, which satisfy the following assumptions: 
\begin{Ass}\label{A1}
Assume that the coefficient $k(t)$ in \eqref{1} satisfies
\begin{align}\label{3}
k(t) = \begin{cases} 
k_{\mathrm{C}} (t) , &  0 \leq |t| < r_0
	, \\ 
 k 
 t^{-2}, & |t| \geq r_0, 
\end{cases}
\end{align}
where  $k_{\mathrm{C}} (t) \in  L^{\infty}
 ((-r_0, r_0))$, $r_0 >0$, is a given large constant and $0 \leq  k <
 m/4$. Moreover, assume that both solutions of
 \eqref{2} with respect to \eqref{3} are included in $C^{1}({\bf R})$ and are
 twice differentiable functions.
\end{Ass} 
Additionally, let $\lambda$, $0 \leq \lambda < 1/2 $, be the smaller one of solutions to $\lambda (\lambda -1)
+ k/m = 0 $, i.e.,  
\begin{align*}
\lambda = \frac{1- \sqrt{1-4k/m}}{2}. 
\end{align*}
Under Assumption \ref{A1}, for $\pm t \geq r_0$, $
( \pm t)^{\lambda}$ and $ (\pm t)^{1- \lambda}$ are linearly independent
solutions to $f''(t) + k t^{-2} f(t)/m =0$. Hence, for $t \geq r_0$, $\zeta _j (t)$ can be represented by $\zeta _1(t) = c_1 t^
{1- \lambda} + c_2 t^{\lambda}$ and $\zeta _2(t) = c_3 t^{1- \lambda} + c_4 t^{\lambda}$, where $x(t) = \zeta _1 (t) x + \zeta _2 (t) p/m = \CAL{O}(t^{1- \lambda})$ and $x'(t)\phi  = m\zeta _1 '(t) x + \zeta _2 '(t) p /m=\CAL{O} (t^{- \lambda})$ holds on $\SCR{S}({\bf R}^n)$. Therefore, 
scattering theory is simpler than the other cases. In contrast, as $x(t) =x + tp/m =\CAL{O}(t)$ and $x'(t) = p/m =\CAL{O}(1)$ holds for $k(t) \equiv 0$, it is expected that the velocity of the classical trajectory of a charged particle is reduced, and the particle decelerated, by the harmonic potentials $k(t)x^2/2$; however, the particle is not trapped. The existence and nonexistence of wave operators for the system (with long-range potentials) were demonstrated in a previous paper by Ishida-Kawamoto \cite{IK}. Hence in this paper, we focus on the characterization of the range of wave operators, i.e., to prove the existence of inverse wave operators. To prove the existence of inverse wave operators, a more rigorous analysis is required than that to prove the existence of wave operators; this is one of the most challenging problems in scattering theory for a time-dependent system. Moreover, it is impossible to decompose the Hamiltonian into the sum of the time-independent free and time-dependent perturbation parts. Scattering theory for such a system has not been considered yet, and our paper provides the first attempt to resolves this issue. Appendix \S{A},  provides an example of $k(t)$ satisfying the assumption \ref{A1}. A more general case can be found in the work by Geluk, Mari\'{c}, and Tomi\'{c} \cite{GMT}; for $k(t)$ that satisfies $ \lim_{t \to \infty}
t^2 k(t) = k$ with $0 \leq k < m/4$, solutions of \eqref{2} satisfy  
\begin{align*} 
\lim_{t \to \infty} \zeta _j (t) /t^{1- \lambda} = \tilde{c}_j   ,
\end{align*}
for $j \in \{1,2\}$, constants $\tilde{c_1 } $ and $\tilde{c}_2 $, and
$\lambda = (1-\sqrt{1-4k/m})/2$.

We define $H(t) = H_0 (t) +V(t)$, where $V(t)$ satisfies
Assumption \ref{A3}. 
\begin{Ass}\label{A3}
$V(t)$ is a multiplication operator with respect to $V(t,x)$, and $V(t,x)
 \in L^{\infty} ({\bf R}; C^{1} ({\bf R} ^n))$ is satisfied if for $\rho_{S , \lambda} > 1/(1-
 \lambda)$,
\begin{align*}
|V(t,x)| \leq C_{S,0} \J{x}^{- \rho _{S , \lambda}}, \quad 
|\nabla V (t,x)| \leq C_{S,1} \J{x}^{- {\rho}_{\mathrm{S} , \lambda}-1 }, \quad
 \J{x} = \sqrt{1 + x^2}
\end{align*}  
holds, where $C_{\mathrm{S}, 0} $ and $C_{\mathrm{S},1}$ are 
 positive constants.
\end{Ass} 

Evidently, under Assumption \ref{A3}, the uniqueness and existence of the unitary propagator for $H(t)$, $U(t,s)$, is guaranteed. By virtue of the simplified $k(t)$, we can decompose propagators $U_0(t,0)$ and $U(t,0) $ into simplified propagators using the approach of Korotyaev \cite{Ko}. 
\begin{prop} \label{PP1}
Let $\SCR{A}$ be $\SCR{A}= x \cdot p + p\cdot x$, and let $U_{S,0} (t,0)$ and $U_S(t,0)$ be propagators for 
\begin{align*}
H_{S,0} (t) := \frac{p^2}{2m |t|^{2\lambda}}, \quad \mbox{and} \quad H_{S} (t) := \frac{p^2}{2m |t|^{2\lambda}} +V(t,|t|^{\lambda} x), 
\end{align*}
then the following factorization of the propagators holds for $\pm t \geq r_0$; 
\begin{align*}
U_0(t,\pm r_0) = e^{im \lambda x^2/(2t)}e^{-i \lambda (\log |t|) \SCR{A}/2} U_{S,0}(t,\pm r_0)
\end{align*}
and 
\begin{align}\label{Pp2}
U(t,\pm r_0) =e^{im \lambda x^2/(2t)}e^{-i \lambda (\log |t|) \SCR{A}/2} U_{S}(t,\pm r_0). 
\end{align}
\end{prop}
By using this proposition, the following can be obtained 
\begin{align*} 
U_0(t,0) = U_0(t, \pm r_0) U_0( \pm r_0,0), \quad U(t,0) = U(t,\pm r_0) U(\pm r_0,0), 
\end{align*}
and therefore
\begin{align*}
U(t,0)^{\ast} U_0(t,0)  &= U( \pm r_0,0)^{\ast} \left( U(t, \pm r_0)^{\ast} 
U_0(t, \pm r_0)
\right) U_0( \pm r_0,0) \\ &= 
 U( \pm r_0,0)^{\ast}  U_S(t, \pm r_0)^{\ast} 
U_{S,0}(t, \pm r_0)  U_0( \pm r_0,0).
\end{align*}
As the usual wave operators, $W^{\pm} $, can be written as 
\begin{align}\label{5}
W^{\pm}  &= \mathrm{s-}\lim_{t \to \pm \infty} U(t,0)^{\ast}U_0(t,0) \\ &= 
 \mathrm{s-}\lim_{t \to \pm \infty} U(\pm r_0,0)^{\ast}  U_S(t,  \pm r_0)^{\ast} 
U_{S,0}(t, \pm r_0)  U_0( \pm r_0,0), \nn
\end{align}
the wave operators $W^{\pm}_S$ for this model are defined as 
\begin{align}\label{S5}
W^{\pm}  _S &= 
 \mathrm{s-}\lim_{t \to \pm \infty}  U_S(t, \pm r_0)^{\ast} 
U_{S,0}(t, \pm r_0)  \\ & =\mathrm{s-} \lim_{t \to \pm \infty} 
U_{S} (t, \pm r_0) e^{\mp i(\pm t)^{1-2 \lambda}p^2/( 2m(1-2 \lambda))}
e^{\pm ir_0 p^2/(2m(1-2 \lambda))}
\nn
.
\end{align}

\begin{thm}\label{T1}
Under Assumptions \ref{A1} and \ref{A3}, $W^{\pm} _S $ exists.
\end{thm}
This Theorem guarantees $\mathrm{Ran} ({W}^{\pm}_S ) \neq \{ 0 \} $. The proof of Proposition \ref{PP1} and Theorem \ref{T1} can be found in \cite{IK} and hence we omit them here. 

By noting this factorization,  we characterize the range of wave operators as
follows: 
\begin{defn} \label{D2}
Let 
$$
 \max \{ 2 \lambda /\rho_{S,\lambda} - \lambda , 0 \}< \rho_{\lambda}< 1- 2 \lambda. 
$$
Define $\CAL{W}
 ^{\pm}_{1,\lambda}$ and $\CAL{W} ^{\pm} _{2,\lambda}$ as follows; 
 \begin{align*}
\CAL{W}_{1, \lambda}^{\pm} &= \left\{ 
\phi  \in L^2 ({\bf R} ^n)\, {\big |} \lim_{t \to \pm \infty}  \left\| (1- \Phi_1
 (p^2))U_S(t, \pm r_0) \phi   \right\|_{L^2({\bf R}^n)} =  0,\ {}^{\exists} \Phi _1 \in C_0^{\infty} ({\bf R} \backslash \{0\}) 
\right\}, \\
\CAL{W}_{2, \lambda}^{\pm} &= \left\{ 
\phi  \in L^2 ({\bf R} ^n)\, {\big |} \lim_{t \to \pm \infty} \int_{|x| \leq |t|^{\rho_{\lambda}}}
 \left| \left( U_S(t, \pm r_0) \phi \right) (x)    \right|^2 dx = 0 
\right\}, 
\end{align*}
and $ {\CAL{W}^{\pm}
 _{\lambda}} := {\CAL{W}^{\pm}_{1,\lambda}} \cap {\CAL{W}^{\pm} _{2,\lambda} }$. 
\end{defn}
The main result is as follows: 
\begin{thm}\label{T2}
 Under Assumption \ref{A1} and \ref{A3}, the characterization of the range of wave operators
 \begin{align*}
\mathrm{Ran} (W^{\pm}_S  ) = \overline{{\CAL{W}}^{\pm}_{\lambda}}
\end{align*} 
holds, where $\overline{{\CAL{W}}^{\pm}_{\lambda}}$ is the closed linear hull of
 $\CAL{W}^{\pm }_{\lambda}$.
\end{thm}
\begin{rem}
Noting \eqref{Pp2}, the range $\CAL{W}_{2, \lambda}^{\pm}$ is equivalent to that 
\begin{align*}
\left\{ 
\phi  \in L^2 ({\bf R} ^n)\, {\big |} \lim_{t \to \pm \infty} \int_{|x| \leq |t|^{\lambda + \rho_{\lambda}}}
 \left| \left( U(t, \pm r_0) \phi \right) (x)    \right|^2 dx = 0 
\right\}.
\end{align*}
Roughly speaking, for $H_0(t)$, we find that $x(t) \sim t^{1- \lambda}$ and $p(t) \sim t^{- \lambda}$, which imply $H_0 (t) \sim t^{-2 \lambda}$. Hence the energy decays  in $t^{-2 \lambda}$. Range $\CAL{W}_{2,\lambda}^{\pm}$ removes the condition so that $ |x(t)| \leq |t|^{\lambda + \rho_{\lambda}} $ on outside of which $ | V(t,x) | \leq C \J{x}^{-\rho_{S}} \leq t^{-(\lambda + \rho_{\lambda}) \rho_S}  < t^{-2 \lambda}$ holds. Hence at least $(\CAL{W}_{2, \lambda}^{\pm})^{c} $ may include all time-dependent bound states. For the case $\lambda = 0 $, more rigorous characterization of range of wave operators due to Kitada-Yajima \cite{KY} has been known; In this range $|x| \leq |t|^{\rho_{\lambda}}$ is replaced by $ |x| \leq n$ and $t$ is replaced by a some sequence $t_n$ with the condition $t_n \to \infty$ as $n \to \infty$.  The aforementioned new range is similar to the one considered by \cite{KY} but which is different. Because of this range, in order to prove Theorem \ref{T2}, we have to prove the scattering particle has velocity faster than at least $\CAL{O}(t^{\rho_{\lambda}-1})$ and which is proven \S{4.2}, moreover  we show that there are no particles scattering with velocity less that $ \ep t^{-2 \lambda}$ if the initial state is included in $\CAL{W}_{\lambda} ^{\pm}$. Thanks to the equivalence of range, our range corresponds to that of \cite{KY} for $\lambda =0$. Since our range employs the continuous limit, it may easy to apply to some advanced studies. 
\end{rem}
Owing to $\CAL{W}_{1, \lambda}^{\pm} $, the positive commutator $\Phi_1 (p^2)i[H_{S,0} , x \cdot p + p \cdot x] \Phi_1 (p^2) $ can be obtained,  which decays over $t^{-2 \lambda}$, and hence $ t^{2 \lambda }\Phi_1 (p^2)i[H_{S,0} , x \cdot p + p \cdot x] \Phi_1 (p^2) $ is positive. In contrast, owing to the non-commutability of $t^{2\lambda}i[ V(t, t^{\lambda} x), \Phi_1 (p^2) ]$, the condition $t^{3 \lambda} | \nabla V (t, t^{\lambda} x) | $ is sufficiently small; however, this fails when $x$ is approximately $0$. By taking $\rho_{\lambda} $ as described above (see \S{2.3} and \S{2.4}), the desired positive commutator can be obtained. 

Before we prove Theorem \ref{T2}, the propagation
estimates for $U_S(t, r_ 0)$ (Proposition \ref{P2}, \ref{P3}, and \ref{P4}) are replaced using the
Mourre type estimates and commutator calculations. This approach is commonly used in proving the
asymptotic completeness of wave operators, not only for quantum systems but
also relativistic equations and quantum field theory (Adachi \cite{A}; Adachi and Ishida \cite{AI}; Adachi and Tamura\cite{AT}; Bachelot \cite{Ba}; Daud\'{e} \cite{Da};
Derezi\'{n}ski and G\'{e}rard \cite{DG}; G\'{e}rard \cite{Ge}; Graf \cite{G};
 and Herbst, M\o ller, Skibsted \cite{HMS}). In particular, we refer to the approach of Graf
\cite{G}. However, as the Hamiltonian $H_S(t)$ never commutes
with the propagator $U_S(t,r_0)$, dealing with time-dependent
energy cut-offs, such as $\varphi_1 (H(t))$, is difficult. 
To overcome this difficulty, we use $\varphi (p^2)$ as the energy cut-off instead of $\varphi (H(t))$ as $\varphi( p^2) $ commutes with $U_{S,0} (t,r_0)$. In contrast, the commutator $i[H_S(t), \varphi (p^2)]$ is not $0$, motivating the
approach of Graf. Hence, in \S{2} we demonstrate that these terms are
included in the $L^1(t;dt)$ class of the suitable cut-off function and 
$\CAL{W}^{\pm}_{\lambda}$. In \S{3}, 
propagation estimates for $U_{S}(t,r_0)$ are provided. In \S{4} the existence of the inverse 
wave operators is proven using
the range of the wave operators $\CAL{W}^{\pm} _{\lambda}$. 
Theorem \ref{T2} is presented in \S{5}. 

Using the reduction scheme of  \cite{Ko}, and in general $k(t)$, the reduced operator is given by
\begin{align*}
H_{S,0} (t) = \frac{p^2}{2m \zeta _1 (t) ^2} + V(t, |\zeta _1 (t) | x)
\end{align*}
and the corresponding propagator $U_{S,0} (t, r_0)$ is $e^{-i p^2 \zeta _2 (t)  /(2m \zeta _1 (t))}e^{i p^2 \zeta _2 (r_0)  /(2m \zeta _1 (r_0))}$, where we use $\zeta _1 (t) \zeta _2 '(t) - \zeta _1 '(t) \zeta _2 (t) =1$. This case contains the following issue; it is difficult to remove the possibility that $\zeta _1 (t) = \CAL{O}(t^{1- \lambda})$ and $\zeta _2 (t) = \CAL{O} (t^{1- \lambda})$ as $t \to \infty$, and $\zeta _2 (t)  / \zeta _1 (t)$ may converge to a constant. For this situation, it is difficult to obtain a propagation estimate, even for $U_{S,0} (t, r_0)$, which is required for the full propagator. The simple solution to this issue is to assume $\zeta _2 (t) = \CAL{O} (t^{1- \lambda})$ and $\zeta _1 (t) = \CAL{O} (t^{\lambda})$. However, for simplified $k(t)$, it is difficult to obtain a $k(t)$ that satisfies this assumption, and only few models have been developed, see Kawamoto-Yoneyama \cite{KaYo}. In our proof, this case is considered by replacing $t^{\lambda }$ by $\zeta _1 (t)$, and $t^{1- 2 \lambda}$ by $\zeta _2 (t)  / \zeta _1 (t)$. In this paper, we only consider the simplified model (Assumption \ref{A1}). However, our proof is a first step in the generalization of $k(t)$.

\section{Auxiliary results}

In this section, we introduce estimations to
simplify the proofs. From this section onwards,
the case  $t> s >0$ with $t \geq 2$ is considered. Other cases can be also proven by the
same approach. Moreover, we also assume that $r_0 >
s >0$ and $r_0 \geq 2$.

\subsection{Notations} 
$C$ is a constant that satisfies $0<C< \infty$ and does
 not depend on any other parameters under consideration. The state
space of a particle is $L^2({\bf R}^n)$. $\left\| \cdot
\right\|$ ($\left\| \cdot \right\|_{L^2({\bf R} ^n)} $ ) is a norm of $L^2({\bf R}
^n)$, and $(\cdot , \cdot)$ is the inner product of $L^2({\bf R}^n)$. The set of bounded operators from $L^2({\bf R} ^n)$ to itself is
given by $\SCR{B}(L^2({\bf R} ^n))$, and components of $\SCR{B} (L^2({\bf R} ^n))$ are denoted by $( \mathrm{bdd} )$. The operator
norm of $\SCR{B}(L^2({\bf R}^n))$, $\left\| \cdot \right\|_{\SCR{B}(L^2({\bf R}
^n))}$, is given by $\left\| \cdot \right\|$. 

Let $f(t)$ be either the $t$-parametrized function or the operator satisfying $|f(t)| \leq C t^{\alpha} $ or $\| f(t) \| \leq C t^{\alpha} $. Then,  $ f (t) \in \CAL{O}(t^{\alpha})$. The
commutator of two operators $A$ and $B$ is $[A, B]$, and 
is defined by 
\begin{align*}
\left( 
i[A, B] \phi, \psi 
\right) := i \left( 
B \phi , A^{\ast} \psi 
\right) - i \left( A \phi , B^{\ast} \psi \right),
\end{align*}
for $\phi, \psi
 \in \D{A} \cap \D{B}$. For all $\psi _0 \in \D{A_0}$ and  for some $b >0$, if $(A_0 \psi_0, \psi_0) \geq b \| \psi _0\| ^2$ holds, $A_0 \geq b$. $A_0$ is positive if $A_0 \geq 0$, and
$\J{x} = \sqrt{1+x^2}$  . 
\subsection{Definitions} 
The cut-off function $F_{\ep} \in C^{\infty} ({\bf R})$ is
defined as 
\begin{align}\label{9}
F_{\ep} (s \leq \theta) = 
\begin{cases}
1, & s \leq \theta - \ep , \\ 
0, & s \geq \theta ,
\end{cases} \quad 
F_{\ep} (s \geq \theta) = 
\begin{cases}
1, & s \geq \theta + \ep,  \\ 
0, & s \leq \theta ,
\end{cases}
\end{align}
where $\ep >0$ is a sufficiently small constant. For $\theta _1
< \theta _2 $, we define
\begin{align*}
F_{\ep} (\theta _1 \leq s \leq \theta _2) = F_{\ep} ( s \geq \theta _1)
 F_{\ep} (s \leq \theta _2).
\end{align*}
Moreover, we define sets $C_{0,\delta}^{\infty} ({\bf R} )$ and
$\SCR{B}_{\delta}^{\infty}({\bf R})$
as follows: 
\begin{align}\label{11}
C_{0,\delta}^{\infty} ({\bf R} ) &= \left\{ 
\phi \in C_0^{\infty} ({\bf R} ) {\big |} 
 \ \mathrm{supp}({\phi}) = \left\{ 
x \in {\bf R} :|x| > \delta
\right\} 
\right\}, \\ 
 \SCR{B}_{\delta}^{\infty} ({\bf R} ) &= \left\{ 
\phi \in \SCR{B}^{\infty} ({\bf R} ) {\big |} 
 \ \mathrm{supp}({\phi}) = \left\{ 
x \in {\bf R} :|x| > \delta
\right\} 
\right\}.
\end{align}
Let $f(t)$ be either a $t$-parametrized function or an operator, which satisfy
\begin{align} \label{ad20-1}
\int_{r_0}^{\infty} |f(t)| dt < \infty, \mbox{ or } \int_{r_0}^{\infty} \|f(t)\| dt < \infty,
\end{align}
then $ f(t) \in  L^1 (t;dt)$. Functions and operators that satisfy \eqref{ad20-1} are denoted by the same notation ($L^1(t)$). For a linear operator $A$, the notation 
$$  (A + (\mathrm{h.c.})) := A + A^{\ast}  $$ is commonly used. For $t-$ parametrized linear operators,
$A(t)$ and $B(t)$, we define the {\em Heisenberg derivative of $A(t)$
associated with $B(t)$} by 
\begin{align*}
{\bf D}_{B(t)} (A(t)) :=  \frac{d}{dt} A(t) + i[B(t),A(t)] . 
\end{align*} 
We also define
\begin{align}\label{theta}
\Theta (t) :=   \frac{x}{|x|}  \cdot \left( 
p- \frac{m(1- 2\lambda)  x}{ t^{1-2 \lambda} }  
\right) .
\end{align}

To simplify the proof, cut-offs are introduced. Functions $\varphi_1$ and $\varphi _2$ are equivalent;
\begin{defn}\label{D1}
Let $\varphi_1 \in C_0^{\infty} ([0, \infty ))$ be defined as follows: 
\begin{align} \label{6}
\varphi_1 (\tau) = 
\begin{cases}
1 & 2 \kappa _1 < \tau < R_1/2, \\ 
0 & \tau < \kappa _1, \ R_1 < \tau,
\end{cases}
\end{align}
where $0 < \kappa_1 \ll 1$ and $R_1 \gg 1$ are constants. Moreover,  we define $\varphi _2
 \in C^{\infty} ([0,\infty)) $ as follows
\begin{align} \label{ad20-7} 
\varphi _2 ' (\tau) \geq 0, \quad 
\begin{cases}
\varphi_2
 (\tau)= 0 & \tau \in [ 0 , \kappa _2], \\ 
\varphi_2 (\tau) = 1 & \tau \in
 (2 \kappa _2 , \infty), 
\end{cases}
\end{align}
where $0 < \kappa _2 \ll 1$ is a constant.
\end{defn}

\subsection{Commutator estimate (Mourre estimates)}

In this subsection, we 
consider the commutator estimate that is closely related to the {\em
Mourre estimate} (see, e.g., Mourre \cite{Mo} and Yokoyama \cite{Yo}). By mimicking the approach of previous works on scattering theory for the Schr\"{o}dinger operator $H=p^2/(2m) +V$, we define the {\em conjugate} operator $\SCR{A}$ as 
\begin{align*}
\SCR{A}  :=   x \cdot p + p \cdot x.
\end{align*}
A straightforward calculation shows that for $t> r_0$
\begin{align} 
i[H_S(t), t^{2 \lambda} \SCR{A}] = {2}{} p^2 /m -2 t^{2 \lambda} (t^{\lambda} x) \cdot \nabla V(t, t^{\lambda} x ).
 \label{14}
\end{align}
To obtain Mourre type estimates for the energy cut-off $\varphi_1 (p^2 )$, the positiveness of the following operator must be obtained:  
\begin{align*}
\varphi _1(p^2) \left( 
{2}{} p^2 /m -2 t^{2 \lambda} (t^{\lambda} x) \cdot \nabla V(t, t^{\lambda} x )
\right) \varphi_1 (p^2) .
\end{align*}
To deduce
the positiveness, $| t^{3 \lambda} x \cdot \nabla V(t,t^{\lambda} x) | \leq C |t|^{-\delta _1}$ for some $\delta _1 >0$. Hence, a specific assumption in the wave
operator range ($\CAL{W}^{\pm}_2 (\lambda )$, see \eqref{Ap1}) is made,  such that $$ |t^{3 \lambda} x \cdot \nabla V(t, t^{\lambda} x) \varphi_2 (x^2 /
t^{2\rho_{\lambda}}) | \leq C t^{-\delta _1} . $$ 
\subsection{Commutator estimates (potential estimates)}
The energy $p^2$ satisfies ${\bf D}_{H_{S,0}(t)} (p^2) = 0$ and ${\bf D}_{H_S(t)} (p^2)  \neq 0$. In the proof of asymptotic completeness, the term ${\bf
D}_{H_S(t)} (p^2) = i[V(t,t^{\lambda} x), p^2]$ appears many
times. Hence, this term is included in the $L^1(t;dt)$
of the suitable cut-off function. 
\begin{lem}\label{L3}
For all $h(\cdot) \in \SCR{B}_{\delta}^{\infty} ({\bf R})$, $\phi \in
 L^2({\bf R}^n)$ and $\rho > 0$,  
\begin{align}
{J} (t) := h(|x|/t^{ \rho }) [V(t,t^{\lambda} x), \varphi _1 (p^2)] \J{t^{\lambda}x}^N \phi = \CAL{O}(t^{- (\rho_{\mathrm{S}, \lambda} +1-N)(\rho + \lambda)
  + \lambda} ) \label{18}
\end{align}
holds, where $N \in \{0,1\}$. 
\end{lem}
\Proof{ 
First, consider the case $N=0$. By the Helffer-Sj\"{o}strand formula (see Helffer-Sj\"{o}strand \cite{HS}), 
\begin{align*}
\varphi_1 (p^2) =\frac{1}{2 \pi i} \int_{\bf C} \overline{\partial
 _{z}} \tilde{\varphi_1} (z) (z- p^2)^{-1} dz d \bar{z},\end{align*}
holds, where $\tilde{\varphi_1}$ is called the {\em almost analytic
extension} of $\varphi_1$. Using $\varphi_1 \in C^{\infty}_0 ({\bf R})$,
$\tilde{\varphi_1}$ is chosen such that $\tilde{\varphi_1} \in
C^{\infty}_0 ({\bf C})$ and 
\begin{align}\label{aaf1}
\left| \overline{\partial _z} \tilde{\varphi}_1(z) \right| \leq C_{M_0} |\mathrm{Im} z|^{M_0} \J{z}^{-M_0 -1}, 
\end{align}
for all $M_0 \in {\bf N}.$ Then 
\begin{align*}
& \left\| 
{h}(|x|/t^{ \rho}) i[V (t,t^{\lambda} x), \varphi_1(p^2)] 
\right\| \leq C \int_{\bf C} |\overline{\partial_z}
 \tilde{\varphi}_1 (z) | \\ & \qquad \times \left\| 
h(|x|/t^{ \rho})(z-p^2)^{-1} \left( 
t^{\lambda} 
p \cdot \nabla V (t,t^{\lambda} x) + (\mathrm{h.c.})
\right) (z-p^2)^{-1}
\right\| dz d\bar{z}
\end{align*} 
holds. 
For sufficiently large $M_1 > (\rho )^{-1} $;
\begin{align*}
 &   {h}(|x|/t^{\rho}) (z- p^2)^{-1}  \\  & = \sum_{M=0}^{M_1-1}
 C_M t^{-\rho M } \left( 
\frac{x}{|x|} \cdot 
p + (\mathrm{bdd}) \right) ^{M }  (z- p^2)^{- M-1 }
 {H}_M (|x|/t^{\rho}) 
 + L^1 (t) ,
\end{align*}
where $(\mathrm{bdd})$ denotes a bounded operator and $H_M \in \SCR{B}_{\delta '}^{\infty}({\bf R})$ with $0< \delta '$. On the support of ${H}_M (|x|/t^{\rho})$, there exists a $\delta' > 0$ such
that $|x| \geq \delta ' t^{\rho}$, yielding 
\begin{align*}
&  \left\| 
{h}(|x|/t^{\rho}) i[V(t,t^{\lambda} x), \varphi_1 (p^2) ] 
\right\| \\ & 
\leq C t^{-(\rho_{\mathrm{S}, \lambda} +1) (\rho + \lambda) + \lambda}\int_{{\bf
 C}} |\overline{\partial _z} \tilde{\varphi} (z)| \J{z}^{(M_1+1)/2}
 |{\mathrm{Im}} 
 z|^{- M_1 -1} dz d \bar{z},
\end{align*}
which combined with \eqref{aaf1} proves \eqref{18} for $N=0$. 
As $(z- p^2)^{-1} \J{t^{\lambda}x} = \J{t^{\lambda}x} (z- p^2)^{-1} +
t^{\lambda} (z-
p^2)^{-1}  \times (\mathrm{bdd})$ and
$ {H}_M (|x|/t^{\rho}) |\nabla V(t,t^{\lambda}x)| \J{t^{\lambda}x} = \ord{t}{- (\rho + \lambda)  \rho
_{\mathrm{S}, \lambda}}$ hold, \eqref{18} is obtained
with $N=1$ for $\rho + \lambda > \lambda$.
}

\section{Propagation estimates}
In this section, we present minimal-velocity
estimates using the Mourre estimate, which play a critical role in determining the existence of inverse wave operators. This approach was considered by
Graf \cite{G} to establish many-body scattering theory. The
approach of Graf for the two-body case is presented in De\'{r}ezinski-G\'{e}rard
\cite{DG}. We provide similar estimates of Propositions 4.4.3, 4.4.4,
and 4.4.7 of \cite{DG} for the Hamiltonian with
time-decaying harmonic potentials. We use a commutator calculation and technical approaches that are introduced in \S{B}.
The following notation is used below; for all $\phi  \in
L^2({\bf R}^n)$, 
\begin{align*}
v(t) = U_S(t,r_0) \phi .
\end{align*}

\subsection{Middle velocity estimate}
 First, To obtain the minimal velocity estimate, the large velocity estimate is required.  
\begin{prop}\label{P2}
Let $\varphi_1$ be that of Definition \ref{D1}. Then for all $\eta _0 >0$ and $\phi \in L^2({\bf R}^n)$, there exists a $C>0$ such that
\begin{align}\label{25}
\int_{r_0}^{\infty} \left\| 
F_{\ep} (\theta \leq |x|/t^{1-2  \lambda} \leq \theta + \eta _0) \varphi
 _1(p^2) v(t)
\right\|^2 \frac{dt}{t} \leq  C \left\| \phi \right\|^2
\end{align}
holds, where $\theta =  2\sqrt{R_1}/(m(1- 2\lambda))$.
\end{prop}
\Proof{ For simplicity, we use the notation 
\begin{align*}
F^{\mathrm{L}} = F_{\ep} (\theta \leq |x|/t^{1-  2\lambda} \leq \theta +
 \eta _0), \quad 
\varphi_1
(p^2) =
\varphi_1 .
\end{align*}
We define $G_{\ep} (t) = \int_{- \infty}^t F_{\ep} (\theta \leq s \leq
\theta + \eta _0 )^2
ds$ and $\theta = 2\sqrt{R_1}/(m(1- 2\lambda)) $, 
then by noting that $\CAL{O} (t^{-2+ 2 \lambda}) \in L^1(t;dt)$ and using Lemma \ref{LVA3}, a straightforward calculation leads to 
\begin{align}
 {\bf D}_{H_S(t)} ( \varphi_1 G_{\ep} (|x|/t^{1- 2\lambda }) \varphi_1) = \frac{\varphi_1 F^{\mathrm{L}}}{mt} \Theta (t)
 F^{\mathrm{L}} \varphi_1 + L^1
 (t). 
\label{e2}
\end{align}
Here, $\Theta (t)$ is the same as that defined in \eqref{theta} with $G_{\ep} (\cdot ) \in \SCR{B}_{\delta}^{\infty} ({\bf
R})$ and Lemma \ref{L3} with $\rho = (1-  2\lambda)$ and $N=0$. By noting the support of $F^{\mathrm{L}}$ and
$\varphi_1$,  
\begin{align*}
\varphi_1 F^{\mathrm{L}} \Theta (t) F^{\mathrm{L}} \varphi_1 = 
- \varphi_1 F^{\mathrm{L}} \left( \frac{m(1-2 \lambda)}{t^{1-2 \lambda}} |x|   + \frac{x}{|x|} p
\right)  F^{\mathrm{L}} \varphi_1. 
\end{align*}
Here, letting $\tilde{\varphi}_1 \in C_0^{\infty} ({\bf R})$ so that $\tilde{\varphi}_1 \varphi_1 = \varphi_1 $ and $\| p \tilde{\varphi} _1 \| \leq \sqrt{ 2R_1} $ holds ( see \eqref{6}). If we consider Lemmas \ref{LVA1} and \ref{LVA3}, then together with Lemma \ref{LVA3}, \eqref{e2} is less than 
\begin{align*}
- \frac{1}{t}
 {\varphi_1} F^{\mathrm{L}}\left( 
\theta (1-2 \lambda) - m^{-1}\sqrt{2R_1}
\right) F^{\mathrm{L}} \varphi_1 + L^1(t), 
\end{align*}
and therefore
\begin{align*}
& \frac{d}{dt} \left( 
 G_{\ep} (|x|/t^{1-2 \lambda }) \varphi_1 v(t), \varphi_1 v(t) \right)  \\ & \leq -\frac{1}{t} 
 \left( 
 \left( 
\theta (1-2 \lambda) - m^{-1}\sqrt{2 R_1}
\right) F^{\mathrm{L}} \varphi_1 v(t),   F^{\mathrm{L}}\varphi_1 v(t)
 \right) 
+ L^1(t). 
\end{align*}
By Lemma \ref{LVA2}, Proposition \ref{P2} can be proven.
}

Second, the middle velocity estimate must be proven.
\begin{prop}\label{P3}
For all $ \phi\in L^2({\bf R}^n)$, $\ep _2 > \ep$, and $\ep _3 > 2 \ep $, there
 exists a $C >0$ such that
\begin{align} 
\int_{r_0}^{\infty} 
\left\| 
\left( 
p - \frac{m(1- 2\lambda)}{t^{1-2 \lambda}} x 
\right) F_{\ep} \left( 
\ep_2 \leq \frac{|x|}{t^{1- 2 \lambda}} \leq \theta + \ep_3
\right)\varphi _1(p^2)v(t)
\right\|^2 \frac{dt}{t} \leq C \left\|
  \phi\right\|^2 \label{d1}
\end{align} 
holds, where $\theta $ is the same as that in Proposition \ref{P2}.
\end{prop}
\Proof{ 
Define $\CAL{R} \in C^{\infty} ({\bf R} ^n)$ as  
\begin{align}\label{28}
\CAL{R}(s) = 
\begin{cases}
s^2, & |s| \geq \ep_2 , \\ 
0, & |s| \leq \ep_2 - \ep, 
\end{cases} \quad 
 \CAL{R}'' \geq 0,
\end{align} 
where $\ep < \ep_2$ and $R'' $ is the Hessian of $(\partial _{jk} R)$. Here, we  assume that $\CAL{R}(s) =0$ holds near the origin to ensure the commutator $i[V(t,t^{\lambda} x), \varphi_1(p^2)]$ is included in $L^1(t;dt)$. However, if we assume  $\CAL{R}(s)=0$ near the origin, \eqref{d1} can be proven by the standard argument since we only consider the one-body problem. Let $\CAL{M} (t)$ be
\begin{align*}
\CAL{M} (t) = \left(m \left( 
p - \frac{m(1- 2\lambda)}{t^{1-2 \lambda}} x
\right) \cdot \nabla \CAL{R} \left( 
\frac{x}{t^{1-  2\lambda}}
\right) + (\mathrm{h.c.}) \right)  + 2m^2 (1-  2\lambda) \CAL{R} \left( 
\frac{x}{t^{1-  2\lambda}}
\right).
\end{align*}
For $\ep_4 = \ep_3 + \ep$, we will prove:
\begin{align*}
&\frac{d}{dt} \left( 
 \CAL{M}(t) F_{\ep} \left( 
\frac{|x|}{t^{1-  2\lambda}} \leq \theta + \ep_4
\right) \varphi _1 v(t) ,  F_{\ep} \left( 
\frac{|x|}{t^{1-  2\lambda}} \leq \theta + \ep_4
\right) \varphi _1 v(t)
\right)  \\ & \quad 
\geq  \left\| F_{\ep} \left( 
\ep_2 \leq \frac{|x|}{t^{1-2 \lambda}} \leq \theta + \ep_3
\right)
\left( p- \frac{m(1-  2\lambda)}{t^{1-2 \lambda}} x \right)  \varphi_1 v(t)
\right\|^2 \frac{1}{t^{}} -  L^1(t).
\end{align*}
By the definition of $\CAL{M}(t)$;
\begin{align*}
\CAL{M} '(t) &= -\frac{m(1-2 \lambda)}{t} \left( \left(
p - \frac{m(1- 2\lambda)}{t^{1-2 \lambda}} x
\right)   {\CAL{R}} '' \left( 
\frac{x}{t^{1-  2\lambda}}
\right)  \frac{x}{t^{1-2 \lambda}} + (\mathrm{h.c.}) \right) 
\end{align*}
and 
\begin{align*}
i[p^2/(2mt^{2 \lambda}), \CAL{M} (t)] &= 
 \frac{1}{t} \left( p - \frac{m(1- 2\lambda)}{t^{1-2 \lambda}} x
\right)   {\CAL{R}} '' \left( 
\frac{x}{t^{1-  2\lambda}}
\right)  p + (\mathrm{h.c.}) + L^1(t), 
\end{align*}
for $t > r_0$. In summary: 
\begin{align} \nn 
{\bf D}_{{H_S}(t)}(\CAL{M} (t))&= \frac{1}{t^{}} \left( 
p- \frac{m(1- 2\lambda)}{t^{1-2 \lambda}} x
\right) {\CAL{R}} '' \left( 
\frac{x}{t^{1- 2 \lambda}}
\right)  \left( 
{p}-  \frac{m(1- 2\lambda)}{t^{1-2 \lambda}} x
\right) \\  & \quad + \Theta _{V} (t;x) + L^1(t), \label{27}
\end{align}
where $\Theta _V (t;x) := i[V (t; t^{\lambda}x), \CAL{M} (t)] \in L^1(t;dt)$. Define $\CAL{L} (t)$ as follows 
\begin{align*}
\CAL{L}(t) = \varphi_1 F_{\ep} \left( 
 \frac{|x|}{t^{1- 2 \lambda}} \leq \theta + \ep_4 
\right) \CAL{M}(t)  F_{\ep} \left( 
\frac{|x|}{t^{1- 2 \lambda}} \leq \theta + \ep_4  
\right) \varphi_1,
\end{align*}
with $\ep_4 = \ep_3 + \ep >0$. For simplicity, we denote 
$$ F_{\ep} \left( 
 {|x|}/{t^{1-  2\lambda}} \leq  \theta + \ep_4 
\right) = F^{\mathrm{MS}}, \quad 
F_{\ep} \left(\ep_2 \leq  |x|/t^{1-2 \lambda} \leq \theta + \ep _3
\right) = F^{\mathrm{M}}
. $$ 
Here we note that on the support of $M(t)$, $|x| > (\ep _2 - \ep) t^{1-2 \lambda}$ holds,
yielding 
$$i[V(t,t^{\lambda} x), \varphi_1] F^{\mathrm{MS}} \CAL{M} (t)
F^{\mathrm{MS}} \varphi_1 \in L^1(t;dt) $$ 
by Lemma \ref{L3} with $\rho =
(1-2 \lambda)$ and $N= 0$. From \eqref{27},
\begin{align*}
 {\bf D}_{H_S(t)} \left(  \CAL{L}(t)
\right)   &=  L^1 (t) + \left( \varphi_1 {\bf D}_{{H}_S(t)} ( F^{\mathrm{MS}}) \CAL{M} (t)
 F^{\mathrm{MS}} \varphi_1 + (\mathrm{h.c.}) \right)  \\ & \quad  + \frac{1}{t^{}} 
\varphi_1  F^{\mathrm{MS}}  \left(p- \frac{m(1- 2\lambda)}{t^{1-2 \lambda}} x \right)
     {\CAL{R} '' \left( 
\frac{x}{t^{1-  2\lambda}}
\right)}
\left(p- \frac{m(1-2 \lambda)}{t^{1-2 \lambda}} x \right) F^{\mathrm{MS}} \varphi _1
\end{align*}
holds. Here, by the definition of ${\CAL{R}}(\cdot)$ and commutator calculation \ref{LVA3}, 
\begin{align*}
&   \varphi _1 F^{\mathrm{MS}}  \left(p- \frac{m(1- 2\lambda)}{t^{1-2 \lambda}} x \right) \CAL{R} '' \left( 
\frac{x}{t^{1-  \lambda}}
\right)
\left(p- \frac{m(1- 2\lambda)}{t^{1-2 \lambda}} x \right) F^{\mathrm{MS}} \varphi _1 \\ & \geq \varphi_1
\left(p- \frac{m(1-  2\lambda)}{t^{1-2 \lambda}} x \right) (F^{\mathrm{M}})^2
\left(p- \frac{m(1-  2\lambda)}{t^{1-2 \lambda}} x \right) \varphi_1 - L^1(t)
\end{align*}
holds. If we obtain 
\begin{align} \label{a5} 
\left|
 \left( 
 \CAL{M} (t) F^{\mathrm{MS}} \varphi_1 v(t) ,  ({\bf
 D}_{{H_S}(t)}(F^{\mathrm{MS}}))^{\ast} \varphi_1 v(t)
\right)
 \right| \in L^1(t;dt), 
\end{align}
then for all $\phi \in L^2({\bf R}^n)$
\begin{align*}
& \frac{d}{dt} \left( \CAL{L}(t)  v(t),v(t)\right)\\ & \geq \left\| F_{\ep} \left( 
\ep_2 \leq \frac{|x|}{t^{1- 2\lambda}} \leq \theta + \ep_3
\right)
\left( p- \frac{m(1-  2\lambda)}{t^{1-2 \lambda}} x \right)  \varphi_1 v(t)
\right\|^2 \frac{1}{t} -  L^1(t;dt). 
\end{align*}
By the same argument in Proposition \ref{P2}, \eqref{d1} is obtained. Hence, we prove \eqref{a5}.

We define $F_{\mathrm{MM}} = F_{\ep} (\theta + \ep _4 - 2 \ep \leq |x| t^{-1+ 2\lambda} \leq \theta + \ep _4 + \ep) $, assuming $\ep _4 > 3 \ep$. A straightforward calculation leads to 
\begin{align*}
\left\| 
(F^{\mathrm{MS}})' (1- F_{\mathrm{MM}}^2) \right\| =0,
\end{align*}
as $t^{-1 +  2\lambda} |x| \leq \theta + \ep _4 - \ep $ or $\theta + \ep _4 \leq t ^{-1 +  2\lambda}|x|$ holds on the support of $1-F_{\mathrm{MM}}^2$. However, if $\eta _1 \geq \theta + \ep_4 + 2 \ep $, for $\ep _ 4 > 3 \ep$ and $\eta _1 \geq \theta + \ep_4 + 2 \ep $, $F_{\mathrm{MM}} F_{\ep} (\theta \leq |x|/t^{1- 2 \lambda} \leq \eta _1) = F_{\mathrm{MM}}$. 
Thus, if $\tilde{\varphi}_1 = \tilde{\varphi _1 }(p^2)$ is an
Operator, such that $\tilde{\varphi_1} \varphi_1 = \varphi_1 $ and
$\tilde{\varphi}_1 \in C_0^{\infty} ({\bf R})$, \eqref{a5} is less than  
\begin{align*}
& Ct^{-1 }\left\| 
\tilde{\varphi}_1 \Theta (t) (F^{\mathrm{MS}})'  
\right\| \left\| 
M(t) F^{\mathrm{MS}} \tilde{\varphi}_1 
\right\| \left\| 
F_{\mathrm{MM}} \varphi_1 v(t)
\right\|^2 + L^1(t)\\ 
& \leq 
Ct^{-1} \left\| 
F_{\ep} (\theta \leq |x|/t^{1- 2\lambda} \leq \eta _1) \varphi_1 v(t)
\right\|^2+ L^1(t)
.
\end{align*}
By Proposition \ref{P2}, \eqref{a5} is integrable with respect to $t$. 
}

\subsection{Minimal velocity estimate}

The self-adjoint operators, $\SCR{A}$ and $\SCR{C}(t)$, are defined  as follows; 
\begin{align}\nn
\SCR{A}  &:=
x \cdot p + p \cdot x, \\ 
\SCR{C}(t) &:=   \varphi_2 (x^2/t^{2 \rho_{\lambda}})\varphi_1 (p^2) 
 {\SCR{A}} \varphi_1
 (p^2)  \varphi_2 (x^2/t^{2 \rho_{\lambda}}), \label{40}
\end{align}
where $\varphi_2 $ is the same as that in Definition \ref{D1}. We 
define 
\begin{align*}
&\CAL{M}_2 (t) =\left( m \frac{x}{|x|} \cdot \left( 
p  - \frac{m(1- 2 \lambda)}{t^{1-2 \lambda}} x 
\right) (F^{\mathrm{S}})' + (\mathrm{h.c.}) \right) + 2m^2 (1-2 \lambda) F^{\mathrm{S}} \\
&F^{\mathrm{S}} = F_{\ep} (|x|/t^{1- 2 \lambda} \leq \ep _5), \quad 
(F^{\mathrm{S}})' =  F_{\ep}' (|x|/t^{1-  2\lambda} \leq \ep _5)
\end{align*}
and 
\begin{align*}
\CAL{L}_2 (t) =  \CAL{M}_2(t) \frac{\SCR{C}(t)}{t^{1- 2 \lambda}}  \CAL{M}_2 (t). 
\end{align*}
Here, we assume that $3 \ep + \ep_2 < \ep_5 <  \kappa _1 /(m(1-2
\lambda) \sqrt{R_1})$. 
Under Assumption \ref{A3}, the following proposition is obtained:
\begin{prop}\label{P4} 
For all $\phi \in L^2({\bf R}^n)$, there exists a $C>0$ such that 
\begin{align} \label{30}
\int_{r_0}^{\infty} \left\| 
F_{\ep}(|x|/t^{1- 2\lambda} \leq \ep_5) \varphi_2 (x^2/t^{2
 \rho_{\lambda}})\varphi_1 (p^2) v(t)
\right\|^2 \frac{dt}{t} \leq C\left\| \phi \right\|^2
\end{align}
holds. 
\end{prop}
\Proof{ We define
\begin{align}
\varphi_1 (p^2)= \varphi _1, \quad \varphi _2 (x^2/t^{2 \rho_{\lambda}}) =
 \varphi_2, \quad  
 \Theta (t) = p \cdot \frac{x}{|x|} - \frac{m(1- 2 \lambda)}{t^{1-2 \lambda}} |x|, \label{as20}
\end{align} 
and
\begin{align*}
\CAL{Q}(t) = \left( \CAL{L}_2(t) v(t), v(t)\right) = \left( \CAL{M}_2 (t) \frac{\SCR{C}(t)}{t^{1-2 \lambda} } \CAL{M}_2 (t) v(t), v(t)\right)
\end{align*}
for the sake of simplicity. We calculate
$d\CAL{Q} (t)/(dt)$. \\ 
{\bf Step I}. Prove 
\begin{align*}
t^{-1 + 2 \lambda}\left( \left( 
{\bf D}_{H_S(t)} (\CAL{M}_2(t)) \SCR{C}(t)   \CAL{M}_2(t) + (\mathrm{h.c.}) \right) v(t), v(t) 
\right) \in L^1(t;dt).
\end{align*}
By the same calculation in the proof of Proposition \ref{P3};
\begin{align*}
& {\bf D} _{{H}_S(t)}(\CAL{M} _2 (t)) = t^{-1 } \Theta (t)
(F^{\mathrm{S}})'' \Theta (t)
+ \theta _V (t;x) + L^1(t), \\ 
& \theta _V(t;x) = i[V(t, x), \CAL{M}_2 (t)] \in L^1(t;dt).
\end{align*} 
Here, we define $F_{\mathrm{M}}^{\mathrm{S}} = F_{\ep} (\ep_5 -2 \ep \leq
|x|/t^{1- 2\lambda} \leq \ep_5 + \ep)$ and obtain
$(F_{\mathrm{M}}^{\mathrm{S}})^2(F^{\mathrm{S}})'' = (F^{\mathrm{S}})''$.
Then, by noting that $\| \SCR{A} F^{\mathrm{S}} \varphi_1\| \leq C t^{1-2 \lambda}$,  
\begin{align}\nn
&t^{-1 + 2 \lambda} |\left( 
 {\bf D}_{{H}(t)}(\CAL{M}_2(t)) \SCR{C}(t) \CAL{M}_2(t) 
 v(t), v(t) 
\right)| \\ \nn &\leq t^{-2+2\lambda} \left| \left( 
\Theta (t) F_{\mathrm{M}}^{\mathrm{S}} (F^{\mathrm{S}})''
 F_{\mathrm{M}}^{\mathrm{S}} \Theta (t) \varphi_2 \varphi_1 \SCR{A} 
 \varphi_1 \varphi_2 \CAL{M}_2 (t) v(t), v(t)
\right) \right|+ L^1(t) \\  
 & \leq 
C t^{-1 }
\left\| 
\Theta (t) F_{\mathrm{M}}^{\mathrm{S}} \varphi_1 v(t)
\right\| ^2 + L^1(t). \label{ac1} 
\end{align}
Here, we use $|x| \leq t^{1-2 \lambda} $ on the support of
$F_{\mathrm{M}}^{\mathrm{S}}$ and $\varphi_2 F_{\mathrm{M}}^{\mathrm{S}}
= F_{\mathrm{M}}^{\mathrm{S}}$ for $t \gg 1$. By Proposition \ref{P3},
with $\ep _5 - 2\ep > \ep _2 + \ep$, \eqref{ac1} is included in $L^1(t;dt)$.
\\ ~~\\ {\bf Step II}.
Consider the term 
\begin{align*}
 {\bf D}_{{H}_S(t)}(t^{-1 + 2 \lambda}\SCR{C}(t)).
\end{align*}
By a straightforward calculation; 
\begin{align*}
 {\bf D}_{{H}_S(t)}(t^{-1 + 2 \lambda}\SCR{C}(t)) = \sum_{j =1}^6 \CAL{J} _j(t), 
\end{align*} 
where 
\begin{align*}
\CAL{J}_1 (t) &= -2 \rho_{\lambda} t^{-2 + 2 \lambda - 2 \rho_{\lambda}}
 x^2\varphi_2' \varphi_1 \SCR{A}  \varphi_1 \varphi_2 + (\mathrm{h.c.}), \\ 
\CAL{J} _2(t) &= - (1- 2 \lambda) t^{-2  + 2 \lambda} \SCR{C} (t), \\ 
\CAL{J} _3 (t) &=  (t^{-1 -2\rho_{\lambda}}/m) \left( \left( 
x \cdot p \varphi _2' + (\mathrm{h.c.})
\right) \varphi_1 \SCR{A}  \varphi_1 \varphi_2 + (\mathrm{h.c.}) \right) , \\
\CAL{J} _4(t) &=  (2/(mt)) \varphi_2 \varphi_1 p^2 \varphi_1 \varphi_2,
 \\ 
\CAL{J} _5(t) &= t^{-1 + 2 \lambda} \varphi_2 i[ V( t,t^{\lambda} x) ,
 \varphi_1 ( p^2)] \SCR{A}  \varphi_1 \varphi_2 + (\mathrm{h.c.}), \\ 
\CAL{J} _6 (t) &= -2 t^{-1 + 2 \lambda} \varphi_2 \varphi_1 \left( t^{\lambda} x  \cdot \nabla V ( t, t^{\lambda} x) \right)\varphi_1 \varphi_2 .
\end{align*}
{\bf Step II-1}. Prove 
\begin{align*}
t^{-1 + 2 \lambda}\left( 
\CAL{M}_2(t) \left( 
\CAL{J}_1(t)  +  \CAL{J}_5 (t) + \CAL{J}_6 (t)
\right) \CAL{M}_2(t) v(t), v(t)
\right) \in L^1(t;dt).
\end{align*}
As $\| \CAL{M}_2(t) \varphi_1\| \leq C$, it is sufficient to prove $\| \CAL{J}_k (t) \| \in L^1(t;dt)$ with $k= 1,5,6$. By the definition of $\varphi_2$, we note that 
\begin{align*}
||x| \varphi_2 '(x^2/t^{2 \rho_{\lambda}})| \leq C t^{\rho _{\lambda}}, 
\end{align*}
holds. Hence, 
\begin{align} \label{z00}
\CAL{J}_1 (t) = \CAL{O} (t^{-2 + 2\lambda + \rho_{\lambda}}) \in L^1(t;dt), 
\end{align}
where $2 \lambda + \rho_{\lambda} < 1$.
Moreover,
\begin{align} 
|(\CAL{J}_6 (t) \phi, \phi )|  \leq Ct^{-1 + 2 \lambda} \left\| 
\J{t^{\lambda}x}^{-\rho_{\mathrm{S, \lambda} } } \varphi_1 \varphi _2 \phi
\right\|  = {\CAL{O}} (t^{-1 + 2\lambda - \rho_{\mathrm{S},\lambda} (\rho_{\lambda} + \lambda)  })
 \in L^1(t;dt) \label{Ap1}
\end{align}
holds for all $\phi \in L^2({\bf R} ^n)$, on the support of $\varphi_2
(x^2 /t^{2 \rho_{\lambda}})$. By 
\begin{align*}
\left\| \CAL{J} _5 (t) \right\|  &\leq C t^{-1 + 2 \lambda} \left\| 
\varphi_2 [V(t,t^{\lambda} x), \varphi_1] \J{t^{\lambda}x}
\right\| \left\| \J{t^{\lambda}x}^{-1} \SCR{A} (t) \varphi_1 \right\|
\end{align*}
and Lemma \ref{L3} with $\rho = \rho_{\lambda}$ and $N=1$, 
\begin{align}
\label{Ap2}  \CAL{J}_5 (t) = \CAL{O} (t^{-1 +2 \lambda } \times t^{
-\rho_{\mathrm{S}, \lambda} (\rho_{\lambda} + \lambda) + \lambda} \times t^{- \lambda} ) \in  L^1 (t;dt).
\end{align} 
{\bf Step II-2}. Prove $\CAL{J}_3 (t) \geq L^1(t;dt)$. \\ 
\begin{align*}
& ( x \cdot p \varphi _2 ' + \varphi_2 ' p \cdot x ) \varphi_1 \SCR{A} \varphi _1 \varphi_2 \\ & = 
\sqrt{\varphi_2 ' \varphi_2} \varphi_1 (\SCR{A})^2 \varphi _1 \sqrt{\varphi_2 ' \varphi_2}  + \CAL{O}(\J{x} ^2 \varphi_2 ') \times \CAL{O} (t^{-\rho_{\lambda}}) + \CAL{O}(\J{x} \varphi_2 ') + \CAL{O} (\J{x}^3 \varphi_2 '') \times \CAL{O}(t^{-2 \rho_{\lambda}})
\end{align*}
yields 
\begin{align} \label{yy7}
\CAL{J} _3 (t) &= L^1 (t) + (mt^{1+2\rho _{\lambda} })^{-1} \sqrt{\varphi_2 ' \varphi_2} \varphi_1
 \SCR{A} ^2 \varphi_1 \sqrt{\varphi_2 \varphi_2 '}  \geq  L^1 (t). 
\end{align} 
We note that  it is not easy to prove that $\CAL{J}_3 (t)$ is included in $L^1(t;dt)$; however, we can easily prove that this term is greater than $L^1(t;dt)$. 
\\  
{\bf Step II-3}. Prove 
 there exist a $\delta _3 >0$ and $\delta _4 >0$ such that 
\begin{align} \nn
& \left( \left( \CAL{J}_2 (t) + \CAL{J}_4 (t)
\right)
\CAL{M}_2 (t)  v(t), \CAL{M}_2(t)  v(t) \right) \\ & \geq \delta _3 \left\| 
F^{\mathrm{S}} \varphi_2 \varphi_1  v(t)
\right\|^2 \frac{1}{t}  - \delta _4 \left\| 
(F^{\mathrm{S}})' \Theta (t) \varphi_2 \varphi_1 v(t)
\right\| ^2 \frac{1}{t} - L^1 (t) \label{ac2}.
\end{align}
Inequalities
\begin{align*}
 \left\| \SCR{A} (t) \varphi_1 F^{\mathrm{S}} \right\| \leq 2 \ep _5 \sqrt{R_1}
 t^{1 - 2 \lambda}, \qquad 
 \CAL{J} _4 (t) \geq (2 \kappa_1/(mt)) \varphi_2 \varphi_1 ^2 \varphi_2,
\end{align*}
yield  
\begin{align*}
& \left(  (\CAL{J} _2 (t) + \CAL{J}_4 (t))
 \CAL{M}_2(t)  v(t), \CAL{M}_2 (t)  v(t)
 , \right)  \\ & \geq 
t^{-1} \left( 
{2 \kappa_1}/{m} - 2(1-2 \lambda) \ep_5 \sqrt{{R}_1}
\right) \left\| 
\CAL{M}_2(t) \varphi_2 \varphi_1  v(t)
\right\|^2 - L^1(t). 
\end{align*}
Here, we take $\ep _5 >0$ sufficiently small, such that $2 \kappa _1 /m
- 2(1-2 \lambda)\ep_5 \sqrt{R_1} >0$, and divide $\CAL{M}_2 (t) = \CAL{M}_3(t) + \CAL{M} _4 (t)$ by
$\CAL{M}_4 (t) = 2m^2(1- 2 \lambda)F_S$ and $\CAL{M}_3 (t) = \CAL{M}_2
(t) - \CAL{M}_4 (t)$. $\kappa_0 $ is defined as a sufficiently small
constant such that $0< \kappa_0 \ll 1$, then $|\CAL{M}_2(t)| ^2 \geq (1- \kappa_0 ^2)
\CAL{M}_4 (t) ^2 + (1- 4\kappa^{-2}_0) \CAL{M}_3 (t) ^2$. Therefore, \eqref{ac2}
can be obtained. \\
{\bf Step III}. Prove \eqref{30}. \\
By \eqref{z00}, \eqref{Ap1}, \eqref{Ap2}, \eqref{yy7}, and \eqref{ac2}, we obtain 
\begin{align} \nn
& \left( \left( {\bf D}_{H_S(t)} (\SCR{C} (t) / t^{1-2 \lambda})
\right) 
\CAL{M}_2 (t)  v(t), \CAL{M}_2(t)  v(t) \right)  \\ & \geq \delta _3 \left\| 
F^{\mathrm{S}} \varphi_2 \varphi_1  v(t)
\right\|^2 \frac{1}{t}  - \delta _4 \left\| 
(F^{\mathrm{S}})' \Theta (t) \varphi_2 \varphi_1 v(t)
\right\| ^2 \frac{1}{t}  - L^1 (t)\label{z02}
\end{align} 
By \eqref{ac1} and \eqref{z02} with Proposition \ref{P3}, we obtain 
\begin{align*}
\frac{d}{dt}\CAL{Q} (t) \geq \delta _3 \left\| 
F^{\mathrm{S}} \varphi_2 \varphi_1 v(t)
\right\|^2 \frac{1}{t} - L^1(t).
\end{align*}
Thus, similar to the proof of Proposition \ref{P2}, we
obtain 
\begin{align}\label{28}
\int _{r_0}^{\infty} \left\| 
F^{\mathrm{S}} \varphi_2 \varphi_1   v(t)
\right\|^2 \frac{dt}{t} \leq C \left\|  \phi  \right\|^2.
\end{align}

}

\section{Existence of inverse wave operators} 

In this section, we demonstrate the existence of inverse wave operators. We address the cut-offs that appear in Definition \ref{D2} (not definition \ref{D1}), which demand very careful and sensitive calculations. First, we show that the minimal velocity decays, which requires the use of the properties of $\CAL{W}^+_{ \lambda}$. 

Let $\Phi _1 \in C_0^{\infty} ({\bf R} \backslash \{ 0\})$ and $\Phi _2 (\cdot) := F_{\ep} (\cdot \geq 1 - \ep )$. Here we note for all $\phi \in L^2({\bf R}^n)$
\begin{align*}
\left\| 
\left( 
1- \Phi_2 (x^2/t^{2 \rho_{\lambda}})
\right) U_S(t, r_0) \phi
\right\|^2  \leq \int_{|x| \leq t^{\rho_{\lambda}}} \left| \left( U_{S} (t, r_0) \phi \right) (x)
\right|^2 dx .
\end{align*} 
Let $\psi _+ \in \CAL{W}^+_{\lambda}$ is arbitrary. We choose $\Phi_j$, $j=1,2$ be such that 
\begin{align*}
\lim_{t \to \infty} \left( 1- \Phi_j \right) U_S(t, r_0) \psi _+ = 0,
\end{align*}
 where 
\begin{align} \label{ad20-4}
\Phi_1 = \Phi_1 (p^2), \qquad \Phi_2 = \Phi_2 (x^2/{t^{2 \rho_{\lambda}}}).
\end{align} 
As $\Phi _1 \in C_0^{\infty} ({\bf R} \backslash \{ 0\})$, then $ 0 < \tilde{\kappa _1} < \tilde{R}_1 $ such that 
\begin{align}\label{ad20-5}
\mathrm{supp}{\Phi_1} = \{ y \in {\bf R} \, | \, \tilde{\kappa}_1 \leq |y| \leq \tilde{R}_1 \}, \quad \mathrm{supp}{\Phi_2} = \{ y \in {\bf R} \, | \,   |y| \geq 1 - \ep \}.
\end{align} 
\subsection{Estimation of the remainder terms}
 In this section, we estimate 
 \begin{align}\label{12}
A_j (t_1,t_2):= \left| \int_{t_1}^{t_2} \left(\Theta (t) {h} (|x|/t^{1- 2\lambda})
 (1- \Phi_j) u(t) , u(t) \right) \frac{dt}{mt} \right|, 
\end{align}  
where $j \in \{ 1,2\}$, ${h} \in C_{0,\delta}^{\infty}({\bf R} )$, $u(t) =
U_S(t,r_0) \psi_+$, $\psi_+ \in \CAL{W}^{+}_{\lambda}$,  
\begin{align*} 
\Theta (t) :=   \frac{x}{|x|}  \cdot \left( 
p- \frac{m(1- 2\lambda)  x}{ t^{1-2 \lambda} }  
\right),
\end{align*}
and $\Phi _1$ and $\Phi_2$ are equivalent to those in \eqref{ad20-4} and \eqref{ad20-5}. In the following Lemma, we shall prove $A_j (r_0, \infty)$ exists. 
\begin{lem}\label{L1}
Let $A_j(t_1,t_2)$ be the same as that in \eqref{12}. Then 
$$
\lim_{t_1,t_2 \to \infty} | A_j (t_1,t_2)  | = 0 
$$ 
holds.  
\end{lem}
\Proof{ 
Define $h(\cdot) \in C_{0,\delta}^{\infty} ({\bf R} )$ and
$\hat{h} (t) = \int_{- \infty}^{t} h(\tau) d \tau$. Here,
$\hat{h} \in \SCR{B}_{ \delta} ^{\infty} ({\bf R})$ and 
\begin{align*}
B_{j} (t) := \left( 
\hat{h}( |x| / t^{1- 2\lambda}) (1- \Phi_j) u(t), u(t)
\right).
\end{align*}
First, we prove $A_1(t_1,t_2) \to 0$ as $t_1,t_2 \to
\infty$. Noting that 
\begin{align*}
{\bf D}_{H_{S}(t)} (\hat{h} (|x| / t^{1- 2\lambda})) = \frac{\Theta (t)}{m
 t^{}} h(|x|/t^{1- 2\lambda}) + \CAL{O}(t^{-2 + 2 \lambda}), 
\end{align*}
$\CAL{O}(t^{-2 + 2 \lambda}) \in L^1(t;dt)$, then 
\begin{align*}
\frac{d}{dt} B_1 (t) & = \frac{1}{mt} \left( \Theta (t)
h\left( 
{|x|}/{t^{1- 2\lambda}} \right)(1- \Phi_1) u(t), u(t)
\right) \\ & + \left( 
\hat{h}\left( 
{|x|}/{t^{1- 2\lambda}}
\right) i\left[ V(t, t^{\lambda} x) , - \Phi_1
\right]u(t),u(t)
\right) + L^1(t).
\end{align*} 
Consequently, 
\begin{align*}
& |B_1 (t_1) - B_1 (t_2) | = \left| 
\int_{t_1}^{t_2} B'_1(t) dt
\right| \\ & \geq \left| 
 \int_{t_1}^{t_2}  
 \left( \Theta (t)
h\left( |x|/t^{1- 2\lambda}
 \right)(1- \Phi_1) u(t), u(t)
\right)
 \frac{dt}{mt} \right| - \int_{t_1}^{t_2} L^1 (t) dt,   
\end{align*}
where we use Lemma \ref{L3} with $\rho = 1-2 \lambda$ and $N=0$.
By the definition of $\Phi_1$ and $\CAL{W}^{+} _{\lambda}$, then $|B_1 (t_1) -B_1 (t_2)|
\to 0$, yielding $A_1 (t_1,t_2) \to 0$.

Second, we prove $|A_2 (t_1,t_2)| \to 0$ as $t_1,t_2 \to \infty$. Here we remark that $\hat{h} (|x|/t^{1- 2 \lambda}) \Phi_2 ' \in L^1(t;dt)$. Then, by using the same approach in the proof of $A_1 (t_1,t_2) \to 0$, we note that it is
sufficient to prove 
\begin{align} \label{13}
C_2(t):= \left| \left( \hat{h}(|x|/t^{1- 2\lambda}) {\bf D}_{H_S(t)} (\Phi_2 (x^2/t^{2 \rho_{\lambda}})) u(t),
 u(t)\right) \right| \in L^1 (t;dt).
\end{align}
However, 
\begin{align*}
\hat{h}(|x|/t^{1- 2\lambda}) {\bf D}_{H_{S}(t)} (\Phi_2) &=
 \frac{\hat{h}(|x|/t^{1-2 \lambda})}{ m t^{2 (\rho_{\lambda}+ \lambda)}}\left( 
\Phi_2 'x \cdot \left( p - \frac{m (1-2\lambda) x}{t^{1-2 \lambda}}\right) + (\mathrm{h.c.})
\right) \\ & \quad +  \frac{2( 1-2\lambda - \rho_{\lambda})\hat{h}(|x|/t^{1-2 \lambda}) \Phi_2 '
 (x^2/t^{2\rho_{\lambda}} ) x^2  }{t^{1+2\rho_{\lambda}}}
\end{align*}
holds. As $|x| \leq 2  t^{\rho_{\lambda}}$ holds on the support of $\Phi_2'$ and
$\rho_{\lambda} = \lambda (1-2 \lambda) < 1-2 \lambda$, we have that for sufficiently large
$t$ and all $r >0$, 
\begin{align*}
| \hat{h}(|x|/t^{1- 2\lambda}) \Phi_2 ' | \leq C t^{- r(1-2 \lambda)} |x|^{r}
 |\Phi_2 '| \leq C t^{-r (1- \lambda -\rho_{\lambda})}.
\end{align*}
This implies 
$$ 
\frac{\hat{h}(|x|/t^{1- \lambda})}{m t^{2(\rho_{\lambda}+ \lambda)}}\left( 
\Phi_2 'x \cdot \left( p - \frac{m (1-2\lambda) x}{t^{1-2 \lambda}}\right) + (\mathrm{h.c.})
\right) \Phi_1 \quad \mbox{and} \quad  \frac{2( \lambda- \rho_{\lambda})\hat{h}\Phi_2 '
 (x^2/t^{2\rho_{\lambda}} ) x^2  }{t^{1 + 2\rho_{\lambda}}}
$$
are included in $L^1(t;dt)$. Hence, $C_2(t)$ can be divided into 
\begin{align*}
&\left( \Lambda (|x|/t^{1-2 \lambda})
\Theta (t) (1- \Phi_1) u(t), u(t)
\right) \frac{1}{mt} + L^1(t), 
\end{align*}
for some $\Lambda (\tau) \in C_{0,\delta}^{\infty} ({\bf R}_{\tau})$. Note that $C_2(t)$ is integrable with respect to $t$ using \eqref{12} with $j=1$.
}
\begin{Cor}\label{Co1}
For all $j,l \in \{ 1,2 \}$ with $j \neq l$ and $l_1, l_2 \in \{0,1\}$, we define 
\begin{align*}
 A_{j,l,l_1,l_2} (t_1,t_2) := \left| \int_{t_1}^{t_2} \left(\Theta (t) {h} (|x|/t^{1- 2\lambda})
 (1- \Phi_j)\Phi_l^{l_1} (1- \Phi_l)^{l_2} u(t) , u(t) \right) \frac{dt}{m t} \right|,
\end{align*}  
then $A_{j,l,l_1,l_2} (t_1, t_2) \to 0 $, as $t_1,t_2 \to \infty$ holds, and also $A_{j,l,l_1,l_2} (r_0 , \infty)$ exists.
\end{Cor}

\subsection{Minimal velocity decays}
We now provide the decay estimate for the minimal velocity, which plays a critical role in this paper. $\psi_+$ is assumed to be arbitrary in $ \CAL{W}_{\lambda} ^+$, and $\tilde{\kappa}_1$ and $\tilde{R}_1$ are set in \eqref{ad20-5}. $\kappa_j$, $j \in \{ 1,2 \}$, and $R_1 $ in \eqref{6} and \eqref{ad20-7}, are assumed to be negligible, so that $\Phi _j (\cdot) = \Phi_j (\cdot) \varphi_j (\cdot) $. $\ep_k$, $k \in \{ 2,3,4,5 \}$, are equivalent to those in \S{3} with associated $\kappa _j$ and $R_1$. 
\begin{thm}\label{mvd}
For all $\psi _+ \in \CAL{W}^{+}_{\lambda}$, there exists $\ep_5 = \ep_5 (\psi _+)$ such that 
\begin{align}\label{29}
\lim_{t \to  \infty} 
\left\| 
F_{\ep} (|x|/t^{1- 2\lambda} \leq \ep _5) U(t,r_0) \psi _+
\right\| = 0
\end{align}
holds.
\end{thm}
\Proof{ 
Let $u_+(t) = U_S(t,r_0) \psi _+$. Then it follows that 
\begin{align*}
&\left\| F^{\mathrm{S}} u_+(t_1) \right\|^2 - \left\| 
F^{\mathrm{S}} u_+(t_2)
\right\|^2 
\\ & = {2} \int_{t_2}^{t_1}  
{\mathrm{Re}} \left( \Theta (t)
(F^{\mathrm{S}})' u_+(t) , F^{\mathrm{S}} u_+(t) \right)
 \frac{dt}{mt}  + \int_{t_2}^{t_1} L^1(t) dt \\ 
& \leq   C(  \CAL{I}_1 )^{1/2} \times(  \CAL{I}_2 )^{1/2} + C(
 \CAL{I}_3 + 
\CAL{I}_4
)  + \int_{t_2}^{t_1} L^1(t) dt,
\end{align*} 
where 
\begin{align*}
\CAL{I}_1 &= 
\int_{t_2}^{t_1} \left\|  
\Theta (t) (F^{\mathrm{S}})' \Phi_1 u_+(t)
\right\|^2 \frac{dt}{t}, \quad  \CAL{I}_2 = 
\int_{t_2}^{t_1} \left\| F^{\mathrm{S}} \Phi_1 \Phi_2 u_+(t)
\right\|^2 \frac{dt}{t}, \\
\CAL{I}_3 &= \left|
\int_{t_2}^{t_1} \left(  
\Theta (t) (F^{\mathrm{S}})' (1- \Phi_1) u_+(t), F^{\mathrm{S}}  u_+(t)
\right) \frac{dt}{mt} \right| , \\ 
\CAL{I} _4 &= \left| \int_{t_2}^{t_1} 
 \left(  
\Theta (t) (F^{\mathrm{S}})' \Phi_1  u_+(t), F^{\mathrm{S}} \Psi_1 u_+(t)
\right) \frac{dt}{mt} \right| , \\ 
\Psi_1 &= (1- \Phi_1) \Phi_2 + \Phi_1 (1- \Phi_2) + (1-
 \Phi_1)(1- \Phi_2). 
\end{align*}
As $\Phi_1 = \varphi_1 \Phi_1$ and $[(F^S)',  \Phi_1 ] $, $[\Theta (t) , \Phi_1] (F^S) '$ decay in $t$, then 
\begin{align*}
\CAL{I}_1 \leq 
\int_{t_2}^{t_1} \left\|  
\Theta (t) (F^{\mathrm{S}})' \varphi_1 u_+(t)
\right\|^2 \frac{dt}{t} + \int_{t_2}^{t_1} L^1 (t) dt
\end{align*} 
holds, and by Proposition \ref{P3}, $\CAL{I}_1 \to 0$ holds. By 
\eqref{28} and the same argument as the above, $\CAL{I}_2 \to 0$ holds. Moreover, by Lemma \ref{L1} and
Corollary \ref{Co1}, $\CAL{I} _3, \CAL{I}_4 \to 0$ holds, and hence the limit
$\lim_{t \to \infty} F^{\mathrm{S}} u_+(t)$ exists, implying $\lim_{t \to \infty}  
 F^{\mathrm{S}} \Phi_1 \Phi_2 u_+(t)$ exists as $F^S (1 -\Phi_j)u_+(t) \to 0$. Owing to the condition $ \CAL{I}_2 < \infty $, 
$\lim_{t \to \infty}  
 F^{\mathrm{S}} \Phi_1 \Phi_2 u_+(t) =0$. By $(1- \Phi_j) u_+(t) \to 0$, we find $\lim_{t \to \infty}  
 F^{\mathrm{S}}  u_+(t) =0$ . }

\subsection{Existence of inverse wave operators}
In this section, we prove the asymptotic completeness of the wave operators. We will prove that for all $\psi _{\pm} \in \CAL{W}_{\lambda} ^{\pm}$
\begin{align*}
W^{\pm}_{S,\mathrm{In}}  :=  \lim_{t \to \pm \infty}
 U_{S,0}(t,\pm r_0) ^{\ast} U_S(t, \pm r_0)  \psi_{\pm}
\end{align*}
exist. 

By the uniformly boundedness of $ U_{S,0}(t, \pm r_0)^{\ast} U_S(t,\pm r_0)  $, it can be proven that there exists $\Phi _1 = \Phi_1 (p^2) \in C_0^{\infty} \left( {\bf R} \backslash \{ 0 \} \right)$ such that
$$ 
 U_{S,0}(t,\pm r_0) ^{\ast} (1- \Phi_1^2) U_S(t, \pm r_0) \psi_{\pm} \to 0,  
 $$
holds. By Theorem \ref{mvd}, we obtain 
\begin{align*}
&\lim_{t \to \pm \infty } U_{S,0}(t, \pm r_0)^{\ast} \Phi_1^2 U_S(t, \pm r_0) \psi_{\pm} 
 \\ & = 
 \lim_{t \to \pm \infty} 
U_{S,0}(t,\pm r_0) ^{\ast} \Phi_1 ^2(1- F^{\mathrm{S}})
 U_S(t, \pm r_0)\psi_{\pm} + 0.
\end{align*}  
Here, $\left\| V(t, t^{\lambda} x) (1- F^{\mathrm{S}})
\right\| \in L^1(t;dt)$ holds and hence the existence of $W_{S, \mathrm{In}}^{\pm}$ can be demonstrated by proving 
\begin{align*}
U_{S,0}(t, \pm r_0)^{\ast} \Phi_1 ^2{\bf D}_{{H}_{S,0}(t)}(1-F^{\mathrm{S}} )
 U_S(t, \pm r_0)\psi_{\pm}  \in L^1(t;dt).
\end{align*} 
However, by the same calculation as that in the proof of Proposition
\ref{P3}, we obtain 
\begin{align*}
& \left\| U_{S,0}(t,\pm r_0)^{\ast} \Phi_1 ^2 {\bf D}_{{H}_{S,0}(t)}(1-F^{\mathrm{S}} )
 U_S(t, \pm r_0) \psi_{\pm} \right\|  \\ 
& \leq \frac{C}{|t|} \left|  \sup_{\left\|v  \right\|=1} \left(\Theta
 (t) \Phi_1 (F^{\mathrm{S}})' U_S(t, \pm r_0) \psi_{\pm}
 ,  \Phi_1 U_{S,0}(t, \pm r_0) v
\right) \right| + L^1(t), 
\end{align*}
where $\Theta (t)$ is the same as that in \eqref{as20}. We let $\tilde{F}_{\mathrm{S}} = F_{\ep} (\ep_5 - 2 \ep \leq |x|/t \leq \ep_5 +
\ep)$, and $\varphi_1$ and $\varphi_2$ are equivalent to those in Definition \eqref{D1} with $\kappa _1$, $\kappa _2$ and $R_1$ satisfying $\varphi_1 \Phi_1 = \Phi_1$ and $\tilde{F}_S \varphi_2 = \tilde{F}_S$ for all $|t| \geq \tilde{r} \gg 1$ with some $\tilde{r} > r_0$. Noting $\tilde{F}_{\mathrm{S}} (F^{\mathrm{S}})' = (F^{\mathrm{S}}) '$, we obtain 
\begin{align*}
& \pm \int_{ \pm \tilde{r}}^{\pm \infty} \left\|  U_{S,0}(t,\pm r_0)^{\ast}  \Phi_1 ^2{\bf
 D}_{{H}_{S,0}(t)}(1-F^{\mathrm{S}} ) U_S(t, \pm r_0) \psi_{\pm} \right\| dt \\ & 
\leq C \left( 
\pm \int_{\pm \tilde{r}}^{\pm \infty} \left\| \Theta (t) (F^{\mathrm{S}}) ' \Phi_1 U_S(t,\pm r_0) \psi_{\pm}
 \right\|^2 \frac{dt}{m|t|}
\right)^{1/2} \\ 
& \times\left( \pm 
\sup_{\left\| v  \right\|=1}\int_{\pm \tilde{r}}^{\pm \infty} \left\|
 \tilde{F}_{\mathrm{S}} \Phi_1  U_{S,0}(t, \pm r_ 0) v \right\|^2
 \frac{dt}{|t|}
\right)^{1/2} \pm \int_{\pm \tilde{r}}^{\pm \infty} L^1(t) dt \\ & 
\leq C \left( 
\pm \int_{\pm \tilde{r}}^{\pm \infty} \left\| \Theta (t) (F^{\mathrm{S}}) ' \varphi_1 U_S(t,\pm r_0) \psi_{\pm}
 \right\|^2 \frac{dt}{m|t|}
\right)^{1/2} \\ 
& \times\left( \pm 
\sup_{\left\| v  \right\|=1}\int_{\pm \tilde{r}}^{\pm \infty} \left\|
 \tilde{F}_{\mathrm{S}} \varphi_2 \varphi_1   U_{S,0}(t, \pm r_ 0) v \right\|^2
 \frac{dt}{|t|}
\right)^{1/2} \pm \int_{\pm \tilde{r}}^{\pm \infty} L^1(t) dt,
\end{align*}
where we use $\tilde{F}_{\mathrm{S}}  (1- \varphi_2) \varphi_1 \in L^1(t;dt)$. By Proposition \ref{P4} and \eqref{28} with
$V \equiv 0$, we can
also prove that all integration terms in the above inequality
are integrable with respect to $t$. By the Cook-Kuroda method, we conclude
that the inverse wave operators $W^{\pm}_{S,\mathrm{In}} $ exist.

\subsection{Range characterization }

Theorem \ref{T2} is  proven in this section. Suppose that $\overline{\phi}_{\pm} \in
\overline{\CAL{W}^{\pm} _{\lambda}}$, and for any $\ep_0 >0$ take
$\phi_{\pm}  \in \CAL{W}^{\pm} _{\lambda}$ such that $\left\| 
\overline{\phi}_{\pm}  - \phi_{\pm} 
\right\| \leq \ep_0 $ holds. Note that there exists a $u_{\pm} 
\in L^2({\bf R}^n)$ such that 
\begin{align*}
  0 & = \lim_{t \to \pm \infty} \left\| 
u_{\pm}  - U_{S,0} (t, \pm r_0)^{\ast} U_S(t, \pm r_0)  \phi_{\pm}  
\right\| \\ &= \left\| 
W^{\pm}_S  u_{\pm} - \phi_{\pm}  + \overline{\phi}_{\pm}  - \overline{\phi}
 _{\pm} 
\right\|  \geq  \left| \left\| 
W^{\pm} _S u_{\pm}  - \overline{\phi}_{\pm} 
\right\| - \ep_0 \right|
\end{align*}
holds. This implies $\overline{\phi}_{\pm}  \in \mathrm{Ran}(W^{\pm}_S) $, i.e.,
$\overline{\CAL{W}^{\pm} _{\lambda} }\subset \mathrm{Ran} (W^{\pm}  _S)$. 

Next, we 
prove $\mathrm{Ran} (W^{\pm}_S ) \subset \overline{\CAL{W}^{\pm} _{\lambda}}$. For all $\psi_{\pm}
\in L^2({\bf R}^n)$, define $ \phi_{\pm} \in \mathrm{Ran} (W^{\pm} _S)$ as 
$\phi_{\pm} = W^{\pm} _S\psi_{\pm} $. First we prove that for all $\hat{\ep } >0 $ there exists $t_0 \gg 1 $ and $\tilde{\varphi}_1 = \tilde{\varphi}_1(p^2) \in C_0^{\infty} ({\bf R} \backslash \{ 0 \})$ such that for all $|t| \geq t_0$ 
\begin{align*}
& \left\| 
(1- \tilde{\varphi}_1  ) U_S(t,\pm r_0) \phi_{\pm}
\right\| \\ & \leq 
\left\| 
(1- \tilde{\varphi}_1) (U_S(t, \pm r_0) \phi_{\pm} - U_{S,0} (t, \pm r_0)\psi_{\pm} )
\right\|  + \left\| 
(1- \tilde{\varphi}_1 ) U_{S,0}(t, \pm r_0) \psi_{\pm}
\right\| \\ & \leq 
C \left\| 
\phi_{\pm} - U_S(t, \pm r_0)^{\ast} U_{S,0}(t, \pm r_0) \psi_{\pm} 
\right\| + \left\| 
(1- \tilde{\varphi}_1) U_{S,0}(t, \pm r_0) \psi_{\pm} 
\right\| \\ &\leq \hat{\ep}
.   
\end{align*}
Evidently, 
\begin{align*}
C \left\| 
\phi_{\pm} - U_S(t,\pm r_0)^{\ast} U_{S,0}(t, \pm r_0) \psi_{\pm} 
\right\| \to 0,  \quad (\leq \hat{\ep} /2), \quad \mbox{as} \quad |t| \to \infty,
\end{align*}
and hence we prove 
\begin{align} \label{LL7}
\|(1- \tilde{\varphi}_1) U_{S,0}(t, \pm r_0) \psi_{\pm}  \|
\leq \hat{\ep} /2 .
\end{align} 
As $\SCR{F}^{-1} \left(  C_0^{\infty} ({\bf R}^n \backslash \{ 0 \}) \right) $ is dense on $L^2({\bf R}^n)$, for all $\hat{\ep}$ there exists $0< \kappa _1 \ll 1$ and $R_1 \gg 1$ such that $\psi_{\pm}^{\kappa _1, R_1 } \in \SCR{F}^{-1} \left(  C_0^{\infty} ({\bf R}^n \backslash \{ 0 \}) \right)  $ with  support $\mathrm{supp} (\SCR{F}[{\psi}_{\pm} ^{\kappa _1 , R_1}](\xi)) = \{ \xi \, | \, 2 \kappa _1 \leq |\xi| \leq  {R}_1  \}$ that satisfies
\begin{align*}
\left\| 
\psi_{\pm} - \psi_{\pm}^{\kappa _1,R_1}
\right\| \leq \hat{\ep} /2.
\end{align*}
Here, $\tilde{\varphi_1}\in C_0^{\infty} ({\bf R}) $ ensuring 
\begin{align*}
\tilde{\varphi}_1(\eta) = 
\begin{cases}
1 & \kappa _1^2 \leq |\eta| \leq 2R_1 ^2, \\ 
0 & |\eta| \leq \kappa _1^2  /4, \ 4R_1^2 \leq |\eta| , 
\end{cases} \qquad 0 \leq \tilde{\varphi}_1 \leq 1.
\end{align*}
Therefore, $\tilde{\varphi}_1 \in C_0^{\infty} ({\bf R} \backslash \{ 0 \})$ and 
\begin{align*}
\left\| 
(1- \tilde{\varphi}_1 (p^2)) U_{S,0}(t, \pm r_0) \psi_\pm
 ^{\kappa _1 , {R}_1} 
\right\| = 0 . 
\end{align*}
Hence we find 
\begin{align*}
\|(1- \tilde{\varphi}_1) U_{S,0}(t, \pm r_0) \psi_{\pm}  \| \leq 0 + \left\| 
\psi_{\pm} - \psi_{\pm}^{\kappa _1,R_1}
\right\| \leq \hat{\ep} /2, 
\end{align*}
which is \eqref{LL7}. Next we prove that for all $\hat{\ep} >0$, there exists $t_0 \gg 1$ such that for all $|t| \geq t_0$, 
\begin{align*}
\left( 
\int_{|x| \leq |t|^{\rho_{\lambda}}}
\left| \left( 
U_{S,0} (t, \pm r_0) \psi _{\pm} 
\right) (x)
\right| ^2 
dx
\right) ^{1/2} \leq \hat{\ep}. 
\end{align*}  
Alternatively, it is suffice to prove 
\begin{align}\label{ad-2020-1}
\left\| 
F_{\ep} (|x|/|t|^{\rho_{\lambda}} \leq 1 + \ep) U_{S,0} (t, \pm r_0) \psi _{\pm}
\right\| \leq \hat{\ep}. 
\end{align}
We let $\psi_{\pm}^{\kappa _1, R_1 }$ be the same one as above. Then l.h.s of \eqref{ad-2020-1} is smaller than 
\begin{align*}
 & \left\| 
\psi_{\pm} - \psi_{\pm}^{\kappa _1,R_1}
\right\| + \left\| 
F_{\ep} (|x|/|t|^{1- \lambda} \leq \kappa _1/m) U_{S,0} (t, \pm r_0) \psi_{\pm}^{\kappa _1,R_1}
\right\| \\ & \quad + 
\left\| \left( 1- 
F_{\ep} (|x|/|t|^{1- \lambda} \leq \kappa _1/m)  \right)
F_{\ep} (|x|/|t|^{\rho_{\lambda}} \leq 1 + \ep) 
\right\| \left\| \psi_{\pm}^{\kappa _1, R_1 } \right\|.
\end{align*}
By a similar calculation of Proposition 2.1 of \cite{IK}, for any fixed $ R_1$ and $ \kappa _1$;
\begin{align}\label{LL1}
\left\| F_{\ep} (|x|/|t|^{1- 2\lambda} \leq \kappa _1/m) U_{S,0}(t, \pm r_0) \psi_\pm
 ^{\hat{\delta}, \hat{R}} \right\|  \to 0 \quad (\leq \hat{\ep}/4),
\end{align}
as $|t| \to \infty$. As
\begin{align} \label{LL2}
& \left\| 
(1- F_{\ep} (|x|/|t|^{1- \lambda} \leq \kappa _1 /m) ) F_{\ep} (|x|/|t|^{\rho_{\lambda}} \leq 1 + \ep) 
\right\|  \left\| \psi_{\pm}^{\kappa _1, R_1 } \right\| \\ \nn & \leq C |{\kappa _1}|^{-N} |t|^{-(1- \lambda) N} \left\| |x|^N F_{\ep} (|x|/|t|^{\rho_{\lambda}} \leq 1 + \ep) 
 \right\|  \left\| \psi_{\pm}^{\kappa _1, R_1 } \right\| \\ & \leq C |\kappa _1|^{-N} (1+ \ep)^N |t|^{N( \rho_{\lambda} -(1- \lambda))} \left\| \psi_{\pm}^{\kappa _1, R_1 } \right\| \nn
  \\ & \leq
 \hat{\ep}/4 , \quad \mbox{as  } |t| \to \pm \infty \nn
\end{align} 
holds; 
\begin{align*}
\left\|  F_{\ep} (|x|/|t|^{\rho_{\lambda}} \leq 1 + \ep)  U_{S,0}(t, \pm r_0) \psi_{\pm} \right\| & \leq
 \| \psi_{\pm} - \psi_{\pm} ^{\hat{\delta},\hat{R}} \| +
 (\mbox{l.h.s of \eqref{LL1}}) + (\mbox{l.h.s. of \eqref{LL2}})  \\ & \leq  \hat{\ep}
\end{align*}
holds. Hence we have $\mathrm{Ran} (W^{\pm}_S ) \subset \CAL{W}^{\pm}
(\lambda ) \subset 
\overline{\CAL{W}^{\pm} (\lambda )}$. Consequently, 
$\mathrm{Ran} (W^{\pm} ) =
\overline{\CAL{W}^{\pm} (\lambda )}$.

\appendix

\section{Example of $k (t)$}
 
An example of $k(t)$ satisfying 
Assumption \ref{A1} is provided here. Let 
\begin{align*}
k (t) = 
\begin{cases}
k_0 & |t| \leq r_0, \\ 
k t^{-2} & |t| > r_0, 
\end{cases} 
\qquad \sqrt{\frac{k_0}{m}} =: \omega _0.
\end{align*}  
Then for $|t| \leq r_0$; 
\begin{align*}
\zeta _1 (t) = \cos (\omega _0 t), \quad 
\zeta _2 (t) = \omega _0^{-1} \sin (\omega _0 t),
\end{align*}
and for $\pm t  > r_0$; 
\begin{align*}
\zeta _1 (t) = c_{1}  |\pm t|^{1- \lambda} + c_{2}
 |\pm t|^{\lambda}, \quad 
\zeta _2 (t,s) = \left( c_{3}   |t|^{1- \lambda} + c_{4} |t|^{\lambda} \right) \mathrm{sign} ({t}).
\end{align*}
Hence, the condition that $\zeta
_j (t,s) \in C^1 ({\bf R})$ are twice differentiable
functions gives 
\begin{align*}
\cos ( \omega _0 r_0) = c_1 r_0 ^{1- \lambda} + c_2 r_0 ^{\lambda},   & \quad  \mp \omega _0 \sin (\omega _0 r_0) =  \pm \left( c_1 (1- \lambda) r_0^{- \lambda} + c_2 \lambda r_0^{\lambda -1}  \right) \\ 
 \pm \omega _0 ^{-1} \sin (\omega _0 r_0) = \pm \left( c_3 r_0^{1- \lambda} + c_4 r_0 ^{\lambda}\right) , &\quad 
 \cos (\omega _0 r_0) = c_3 (1- \lambda) r_0^{- \lambda} + c_4 \lambda r_0 ^{\lambda -1}.
\end{align*}
This implies 
\begin{align*}
& r_0^{1- \lambda} (1-2 \lambda) c_{1} = - \lambda \cos(\omega _0
 r_0) - r_0 \omega _0 \sin (\omega _0 r_0 ), \\
& r_0^{\lambda} (1-2 \lambda) c_{2} = {(1- \lambda) \cos (\omega _0 r_0 ) + r_0
 \omega _0 \sin (\omega _0 r_0 )}, \\
& \omega _0 r_0^{1- \lambda} (1-2 \lambda) c_{3} = 
\omega _0r_0 \cos (\omega _0 r_0 ) - \lambda \sin (\omega _0
 r_0 ), \\
& \omega _0 r_0^{\lambda} (1-2 \lambda) c_{4}  = - \omega _0 r_0 \cos (\omega _0
 r_0 ) + (1- \lambda) \sin (\omega _0 r_0 ).
\end{align*}  
For all given $\omega_0$, $k_0$, and $k$, by taking $c_{j} $
($j = 1,2,3,4$) as above with $\zeta _j (t)$, then the solution to \eqref{2}
satisfy Assumption \ref{A1}. Here, $k_0 = 0$
because $\omega _0 ^{-1} \sin (\omega _0 r_0) \to r_0 $ as
$\omega _0 \to 0$.

\section{Operator calculations}
In this section, \eqref{29} is obtained using operator calculations. Techniques of the operator calculation are used here, and are summarized below. 
\begin{lem}\label{LVA1}
For some self-adjoint operator $A_0$, bounded operator $B_0$, and for all $\psi_0 \in \D{A_0}$, 
\begin{align*}
\left( 
B_0 A_0 \psi _0, A_0 \psi _0
\right) \leq \| B_0 \|  \| A _0\psi _0\|_2^2
\end{align*} 
holds.
\end{lem}

The following useful Lemmas are introduced to provide the propagation estimates; 
\begin{lem} \label{LVA2}
Let $G_0(t)$ be a uniformly bounded operator for every $t$ and self-adjoint operator for every fixed $t$. Let $W_0(t,s)$ be a unitary time-evolution propagator. Then, for every $\psi_0 (s)\in L^2({\bf R}^n)$, if the inequality 
\begin{align*}
& \frac{d}{dt} \left( 
G_0(t) W_0(t,s) \psi_0(s), W_0(t,s) \psi_0 (s)
\right)  \\ & \qquad \geq c_0 \left( 
A_0(t) ^{\ast}A_0(t) W_0(t,s) \psi_0(s), W_0(t,s) \psi_0 (s)
\right)  - C |t|^{-1- \ep_0} \| \psi_0 (s) \|_2^2 
\end{align*}
holds for some $c_0 >0$, $\ep _0 >0$, and bounded operator $A_0(t)$, 
\begin{align*}
\int_1^{\infty} \left\| 
A_0(t) W_0(t,s) \psi_0 (s)
\right\| _2^2 dt \leq C \| \psi _0 (s)\|_2^2
\end{align*}
holds.
\end{lem} 
This proof can be found in the literature, e.g., \cite{DG} Lemma B.4.1. To calculate the commutator between the self-adjoint operator-valued function and some self-adjoint operator, the following commutator expansions are used (see, e.g. \cite{SS} or Lemma C.3.1 of \cite{DG});   
\begin{lem}[Commutator expansions]\label{LVA3}
Let $f$ be included in $C^1({\bf R})$, $f'$ be bounded on ${\bf R}$, and $\rho > 0$. Then 
\begin{align*}
[f(|x| /t^{\rho}), \varphi_1 (p^2) ] = \CAL{O}(t^{-\rho})
\end{align*}
holds.
\end{lem}
This lemma can be proven easily by applying the Helffer-Sj\"{o}strand formula to $\varphi _1$ and performing the same calculation as that in the proof of Lemma \ref{L3}.

\section{Time-decaying magnetic field}
Let us consider the Hamiltonian 
\begin{align*}
H_{0,B} (t) = \frac{1}{2m} \left(p_1 + \frac{qB(t)}{2} x_2 \right)^2 +   \frac{1}{2m} \left(p_2 - \frac{qB(t)}{2} x_1 \right)^2, \quad 
\mbox{on  } L^2({\bf R}^2), 
\end{align*}
where $q \neq 0$ is the charge of a charged particle and $B(t) \in L^{\infty} ({\bf R})$ is the intensity of the time-dependent magnetic field 
${\bf B} (t) = (0,0, B(t))$. We define $U_{0,B} (t,s)$ as the propagators for $H_{0,B} (t)$. Note that
\begin{align*}
H_{0,B} (t) = \frac{1}{2m} p_{(2)}  ^2 +  \frac{q^2B(t)^2}{8m}  x_{(2)}^2 - \frac{qB(t)}{2m}L_{(2)}, 
\end{align*}
where $x_{(2)} = (x_1,x_2)$, $p _{(2)}= (p_1,p_2)$, and $L_{(2)}= x_1 p_2- x_2 p_1$; $L_{(2)}$ is the angular momentum. Moreover, define $H_B(t) = H_{0,B} (t) + V(t)$ where $V(t)$ satisfies Assumption \ref{A3}. Let $H_{OS,0} (t)$ and $H_{OS} (t)$ be defined as 
\begin{align*}
H _{OS} (t) := H_{OS,0} (t) + \hat{V}(t) := \frac{p_{(2)}^2 }{2m} + \frac{q^2B(t)^2}{8m} x_{(2)} ^2 + \hat{V}(t), 
\end{align*} 
where $\hat{V}(t)$ is a multiplication operator of $V(t, \hat{x}(t))$ with 
\begin{align*}
\hat{x}(t) := \MAT{\cos (\Omega (t) /2) &  \sin (\Omega (t) /2) \\ - \sin (\Omega (t) /2) & \cos(\Omega (t)/2)} 
\MAT{x_1 \\ x_2 }, \quad \Omega (t) = \int_0^t \frac{qB(\tau)}{m} d\tau, 
\end{align*}
see \cite{AK2}. Let $U_{B} (t,s)$, $U_{OS,0}$, and $U_{OS} (t,s)$ be propagators for $H_{B} (t)$, $H_{OS,0} (t)$, and $H_{OS} (t)$, respectively. Then,
\begin{align*}
U_{B} (t,0) = e^{i \Omega (t) L_{(2)} /2} U_{OS} (t,0).
\end{align*} 
Hence, the wave operators for this system $\CAL{W} ^{\pm}_{B} (s)$ are defined as 
\begin{align*}
{W} ^{\pm}_{B} = \mathrm{s-} \lim_{t \to \pm \infty} 
U_{B}(t,0)^{\ast} U_{0,B} (t,0) =  \mathrm{s-} \lim_{t \to \pm \infty} 
U_{OS}(t,0)^{\ast} U_{OS,0} (t,0).
\end{align*}
Hence, rewriting 
\begin{align*}
\frac{q^2 B^2(t)}{4m} = k(t)
\end{align*}
and assuming that $k(t)$ satisfies Assumption \ref{A1}, we prove the existence and characterization of range of wave operators for the magnetic field case.


\begin{thebibliography}{99}

\bibitem{A} Adachi, T.: Propagation estimates for N-body Stark
	Hamiltonians, Ann. del'I.H.P. \textbf{62}, 409-428, (1995).

\bibitem{AI} Adachi, T., Ishida, A.: Scattering in an external electric field asymptotically constant in time, J. Math. Phys. \textbf{52}, (2011).


\bibitem{AK2} Adachi, T., Kawamoto, M.: Quantum scattering in a
	periodically pulsed magnetic field,
	Ann. H. Poincar\'{e} \textbf{17}, 2409-2438, (2016). 


\bibitem{AT} Adachi, T., Tamura, H.: Asymptotic completeness for
	long-range many-particle systems with Stark
	effect, J. Math. Soc. Univ. Tokyo \textbf{2}, 77-116, (1995). 

\bibitem{Ba} Bachelot, A.: Gravitational scattering of electromagnetic
	field by Schwarzschild
	black-hole, Ann. del'l. H. P. \textbf{54}, 261-320, (1991).

\bibitem{Da} Daud\'{e} T.: Time-dependent scattering theory for charged Dirac fields on a Reissner-Nordstr\"{o}m black hole,  J. Math. Phys. \textbf{51}, (2010).  

\bibitem{DG} Derezi\'{n}ski, J., G\'{e}rard, C.: Scattering theory
	    of classical and quantum N-particle systems, Text
	Monographs. Phys. Springer, Berlin, (1997).

\bibitem{GMT} Geluk, J. L., Mari\'{c}, V., Tomi\'{c}, M.: On regularly varying solutions of second
order linear differential equations, Differential and Integral Equ.
\textbf{6}, 329-336, (1993).

\bibitem{Ge} Gerard, C.: Scattering theory for Klein-Gordon equations
	with non-positive energy, Ann. Henri. Poincar\'{e} \textbf{13},
	883-941, (2012). 

\bibitem{G} Graf, G-, M.: Asymptotic completeness for N-body short-range quantum system: A new proof, Comm. Math. Phys. \textbf{132}, 73-101, (1990). 

\bibitem{HLS} Hagedorn, G. A., Loss, M., and Slawny, J.: Nonstochasticity of time-dependent
quadratic Hamiltonians and the spectra of canonical transformations,
J. Phys. A \textbf{19}, 521-531 (1986).

\bibitem{HMS} Herbst, I., M\o ller, J.S., Skibsted, E.: Asymptotic
	completeness for N-body Stark Hamiltonians, Comm. Math. Phys.
	\textbf{174}, 509-535, (1996).  

\bibitem{HS} Helffer, B., Sj\"{o}strand, J.: Equation de
	    Schr\"{o}dinger avec champ magn\'{e}tique et equation de
	    Harper, Springer Lecture Notes in Physics \textbf{345}, 118-197, (1989). 

\bibitem{Hu} Huang, M. J.: On stability for time-periodic perturbations of harmonic oscillators,
Ann. Inst. H. Poincar\'{e} Phys. Th\'{e}or. \textbf{50}, 229-238 (1989).

\bibitem{Ho} Hochstadt, H.: Functiontheoretic properties of the discriminant of Hill's equation, Math. Z. \textbf{82}, 237-242, (1963).

\bibitem{IK} Ishida, A., Kawamoto, M.: Existence and nonexistence of wave operators for time-decaying harmonic oscillators, to appear

\bibitem{Ka} Kawamoto, M.: Mourre theory for time-periodic magnetic fields, J. Funct. Anal., \textbf{277}, 1-30, (2019). 

\bibitem{KaYo} Kawamoto, M., Yoneyama, T.: Strichartz estimates for harmonic potential with time-decaying coefficient, J. Evol. Eqn., \textbf{18}, 127--142, (2018).  

\bibitem{Kyo} Kato, K., Yoneyama, T.: Characterization of the ranges of wave operators for Schr\"{o}dinger equations with time-dependent short-range potentials via wave packet transform, to appear in Funkcialaj Ekvacioj, \textbf{63}, 19-37, (2020). 

\bibitem{KY} Kitada, H., Yajima, K.: A scattering theory for
	    time-dependent long-range potentials, Duke Math. J.
	    \textbf{49}, 341-376, (1982).

\bibitem{Ko} Korotyaev, E. L.: On scattering in an external,
	    homogeneous, time-periodic magnetic field,
	    Math. USSR-Sb. \textbf{66}, 499-522, (1990).



\bibitem{Mo} Mourre, E.: Absence of singular continuous spectrum for
	    certain selfadjoint operators, Comm. Math. Phys. \textbf{91}, 391-408, (1981).

\bibitem{Na} Naito, M.: Asymptotic behavior of solutions of second
	    order differential equations with integrable coefficients,
	    Trans. A.M.S. \textbf{282}, 577-588, (1984). 
	    

\bibitem{SS} Sigal, I. M., Soffer, A.: The $N$-particle scattering problem: asymptotic completeness for short-range systems, Ann. Math. \textbf{126}, 35-108, (1987).  

\bibitem{W} Willett, D.: On the oscillatory behavior of the solutions of second order linear differential equations,
Ann. Polon. Math. \textbf{21}, 175-194, (1969).

\bibitem{Yaf} Yafaev, D. R.: Scattering subspaces and asymptotic completeness for the time-dependent Schr\"{o}dinger
equation, Mat. Sb. \textbf{118} (160), 262-279, (1982). 



\bibitem{Yo} Yokoyama, K.: Mourre theory for time-periodic systems,
	    Nagoya Math. J. \textbf{149}, 193-210, (1998).

\end{thebibliography}
\end{document}